\documentclass{WileyMSP-template}
\usepackage{amssymb}
\graphicspath{ {./images/} }

\begin{document}

\pagestyle{fancy}

\title{Coherent Ising Machines with Optical Error Correction Circuits}

\maketitle


\author{Sam Reifenstein}
\author{Satoshi Kako}
\author{Farad Khoyratee}
\author{Timoth\'ee Leleu}
\author{Yoshihisa Yamamoto*}


\dedication{}

\begin{affiliations}
Sam Reifenstien, Satoshi Kako, Farad Khoyratee, Yoshihisa Yamamoto\\
PHI (Physics \& Informatics) Laboratories, NTT Research Inc.\\
940 Stewart Drive, Sunnyvale, CA 94085, U.S.A.\\
Email Address:yoshihisa.yamamoto@ntt-research.com\\

Timoth\'ee Leleu\\
International Research Center for Neurointelligence, The University of Tokyo,\\
7-3-1 Hongo Bunkyo-ku, Tokyo 113-0033, JAPAN\\

\end{affiliations}



\begin{abstract}

We propose a network of open-dissipative quantum oscillators with optical error correction circuits. 
In the proposed network, the squeezed/anti-squeezed vacuum states of the constituent optical parametric oscillators below the threshold establish quantum correlations through optical mutual coupling, while collective symmetry breaking is induced above the threshold as a decision-making process. This initial search process is followed by a chaotic solution search step facilitated by the optical error correction feedback.
As an optical hardware technology, the proposed coherent Ising machine (CIM) has several unique features, such as programmable all-to-all Ising
coupling in the optical domain, directional coupling ($J_{ij} \neq J_{ji}$) induced chaotic behavior, and low power operation at room temperature. 
We study the performance of the proposed CIMs and investigate how the performance scales with different problem sizes. 
The quantum theory of the proposed CIMs can be used as a heuristic algorithm and efficiently implemented on existing digital platforms. 
This particular algorithm is derived from the truncated Wigner stochastic differential equation.
We find that the various CIMs discussed are effective at solving many problem types, however the optimal algorithm is different depending on the instance.
We also find that the proposed optical implementations have the potential for low energy consumption when implemented optically on a thin film {\textmd{LiNbO$_3$}} platform.

\end{abstract}


\section{Introduction}
Combinatorial optimization problems are ubiquitous in modern science, engineering, medicine, and business. 
Such problems are often NP-hard; hence, their runtime on classical digital computers is expected to scale exponentially. 
A representative example of NP-hard combinatorial optimization problems is the non-planar Ising model.{\scriptsize $^{\cite{Barahone1982}}$} 
Special-purpose quantum hardware devices have been developed for finding
solutions of Ising problems more efficiently compared tothan standard heuristic approaches.
For example, a quantum annealing (QA) device exploits the adiabatic evolution of pure-state vectors using a time-dependent Hamiltonian.{\scriptsize $^{\cite{Johnson2011, Boxio2014}}$} 
Another example is a coherent Ising machine (CIM), which exploits the quantum-to-classical transition of mixed-state density operators in a quantum oscillator network.{\scriptsize $^{\cite{Marandi2014, Inagaki2016NP, McMahon2016, Inagaki2016}}$} 
Performance comparisons between QA devices and CIMs for various Ising models, such as complete,
dense, and sparse graphs, have been reported.{\scriptsize $^{\cite{Hamerly2019}}$} 
Furthermore, theoretical performance comparisons between ideal gate-model quantum computers, implementing either Grover’s search algorithm or the adiabatic quantum computing algorithm, and CIMs have been reported recently.{\scriptsize $^{\cite{Sankar2021}}$} 
Although CIMs with all-to-all coupling among spins are highly effective, the use of an external FPGA circuit as
well as an analog-to-digital converter (ADC) and a digital-to-analog converter (DAC) not only results in considerable energy dissipation but also introduces a potential bottleneck for high-speed operation.

\medskip
The standard linear coupling scheme of CIMs has been found to suffer from amplitude heterogeneity among the constituent quantum oscillators. Consequently, the Ising Hamiltonian is incorrectly mapped to the network loss, resulting in unsuccessful operation, especially in frustrated spin systems.{\scriptsize $^{\cite{Wang2013}}$} 
A novel error-correcting feedback scheme has been developed to resolve this problem{\scriptsize $^{\cite{Leleu2019, Leleu2021}}$}, which makes the solution accuracy of CIMs comparable to that of state-of-the-art heuristics such as break-out local search (BLS).
{\scriptsize $^{\cite{BLS}}$} In this paper, we introduce a novel CIM architecture in which the error correction is implemented optically. In the proposed architecture, computationally intensive matrix–vector multiplication (MVM) and a nonlinear feedback function are implemented using phase-sensitive (degenerate) optical parametric amplifiers, which are essentially the same device as the main-cavity optical parametric oscillator (OPO). 
This new CIM architecture can potentially be implemented monolithically in future photonic integrated circuits using thin-film {\textmd{LiNbO$_3$} platforms. {\textmd{LiNbO$_3$} platforms.{\scriptsize $^{\cite{TFLN}}$}

\medskip
A network of open dissipative quantum oscillators with optical error correction circuits is promising not only as a future hardware platform but also as a quantum-inspired algorithm because of its simple and efficient theoretical description.
Numerical simulation of the time evolution of an N-qubit quantum system requires $2^N$ complex-number amplitudes. However, for a quantum oscillator network, various phase-space techniques of quantum optics have been developed over the last four decades.{\scriptsize $^{\cite{Drummond1980, Drummond1981, Walls2007}}$} 
The complete description of a network of quantum oscillators is now possible using $N$ (or $2N$) sets of stochastic differential equations (SDEs) based on positive-P,{\scriptsize $^{\cite{Takata2015}}$} truncated Wigner {\scriptsize $^{\cite{Maruo2016, Inui2020, Inui2021}}$} or truncated Husimi {\scriptsize $^{\cite{Inui2020, Inui2021}}$} representations of the master equations. 
These SDEs can be used as heuristic algorithms on modern digital platforms. To completely described a network of low-Q quantum oscillators, a discrete map technique using a Gaussian quantum model is available, which is also computationally efficient.{\scriptsize $^{\cite{Ng2021}}$} 

\medskip
Similarly, a network of dissipation-less quantum oscillators with adiabatic Hamiltonian modulation is described using a set of N deterministic equations, which can also be used as a heuristic algorithm on modern digital platform.{\scriptsize $^{\cite{Goto2019, Tatsumura2021, Goto2021}}$}  
Such heuristic algorithms are called simulated bifurcation machines (SBMs),{\scriptsize $^{\cite{Goto2019, Goto2021}}$} a variant of which will be studied in Section 6.
Although the original SBM is inspired by dissipation-less adiabatic quantum computation, the version of SBM discussed in this paper (dSBM) is not a true unitary system, 
as dissipation is artificially added using inelastic walls to improve the performance of the algorithm. 
As both algorithms involve MVM as a computational bottleneck when simulated on a digital computer, we use the number of MVMs as the metric for performance comparison.
We find that both types of systems have very similar performance in most cases, except graph types with great variation in vertex degree, where the SBM struggles consistently.

\section{Semi-classical Model for Error Correction Feedback}

In this section, we will describe several mutual coupling and error correction feedback schemes for CIMs. To simplify our argument, we consider a semi-classical deterministic picture. {\scriptsize $^{\cite{Wang2013}}$}
The semi-classical model treated
in this section is an approximate theory for the following fictitious machine. At an initial time t = 0,
each signal pulse field is prepared in a vacuum state (Figure \ref{fig:semi-classical}(a)), and each error pulse field is prepared in a weak coherent state (Figure \ref{fig:semi-classical}(b)). When the pump fields $p$ and $p_i$ are
switched on at $t \geq 0$, a vacuum field incident on the extraction beam splitter BS$_e$ from an open port is
squeezed/anti-squeezed by a phase-sensitive amplifier (PSA) in this optical delay line (ODL) CIM, as shown in Figure \ref{fig:semi-classical}(c).
In other words, the vacuum fluctuation in the in-phase component $\tilde{X} = \frac{1}{2}\left(\hat{a} + \hat{a}^{\dagger}\right)$
is deamplified by a factor of
1/G, while the vacuum 
fluctuation in the quadrature-phase component $\hat{P} = \frac{1}{2i}
\left(\hat{a} - \hat{a}^{\dagger}\right)$
is amplified by a factor of G. Similarly, the vacuum fluctuations incident on the OPO pulse field owing to any linear loss
of the cavity are all squeezed by the respective PSA. Moreover, the pump field and feedback injection field fluctuations along the in-phase component are also deamplified by the respective PSA (Figure \label{fig:semi-classical}(c)).

\begin{figure}[!htb]
	\centering
	\includegraphics[scale=0.1]{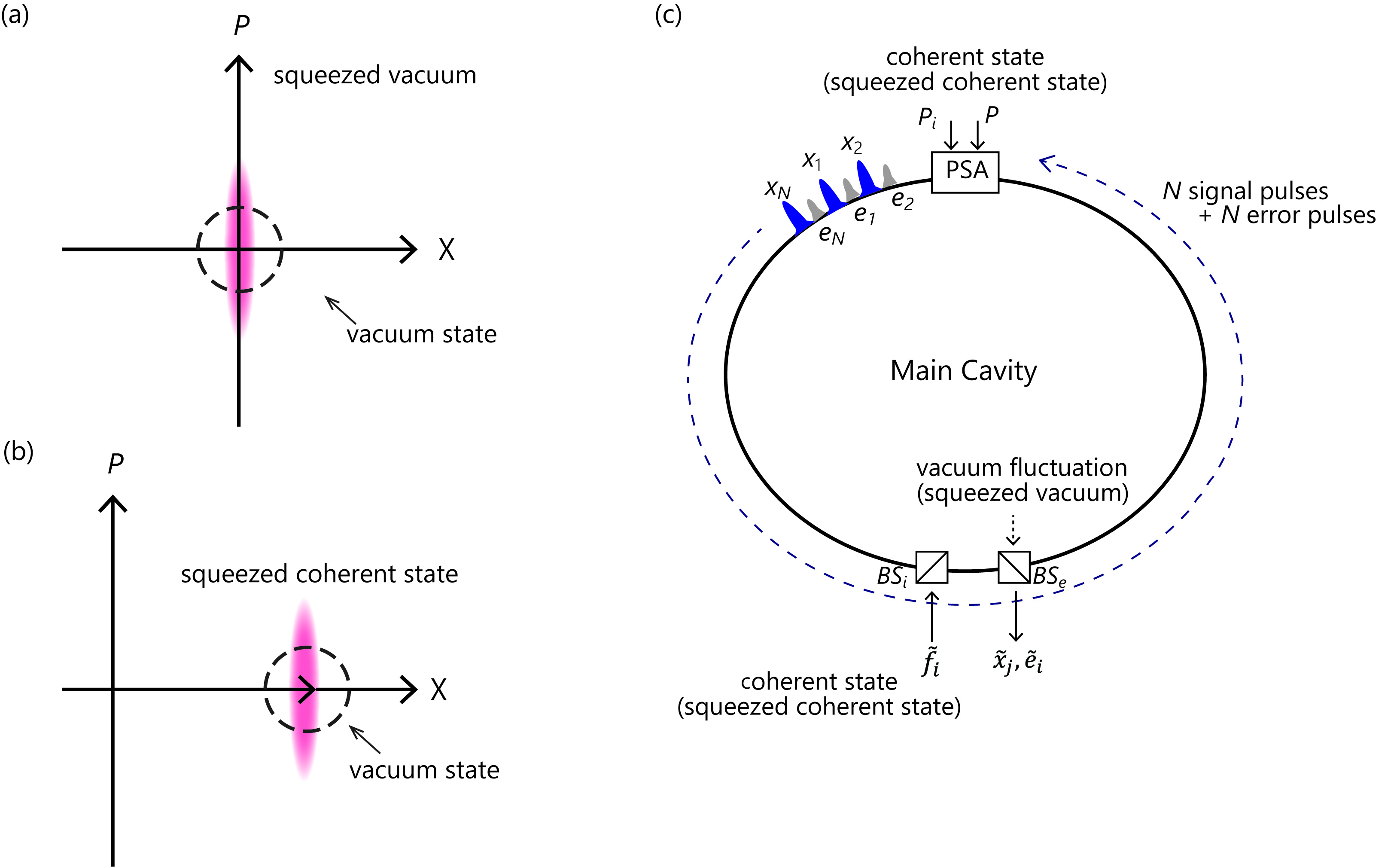}
	\caption{(a) Vacuum state and squeezed vacuum state. (b) Coherent state and squeezed coherent state. (c) Conventional CIM with vacuum noise injected from reservoirs, and a modified CIM with suppressed reservoir noise.}
	\label{fig:semi-classical}
\end{figure}

\medskip
The truncated Wigner stochastic differential equation (W-SDE) for such a quantum-optic CIM with squeezed reservoirs has been derived and studied previously.{\scriptsize $^{\cite{Kako2020}}$} This particular CIM achieves the maximum quantum correlation
among OPO pulse fields along the in-phase component as well as the maximum success probability,{\scriptsize $^{\cite{Kako2020}}$}
because the quantum correlation among OPO pulse fields is formed by the mutual coupling of the vacuum fluctuations of OPO pulse fields without the injection of uncorrelated fresh reservoir noise in such a system. 
The following semi-classical model is considered as an approximate theory of the above-mentioned W-SDE in the limit of a large deamplification factor $(G \gg 1)$. 
A full quantum description of a more realistic CIM with optical error correction circuits (without reservoir engineering) is given in Section 5.
\medskip
To overcome the problem of amplitude heterogeneity in the CIM {\scriptsize $^{\cite{Wang2013}}$}, the addition of an auxiliary variable for error detection and correction has been proposed.{\scriptsize $^{\cite{Leleu2019,Leleu2021}}$}
This system has been studied as a modification of the measurement feedback CIM.{\scriptsize $^{\cite{Kako2020}}$} 
The spin variable (signal pulse amplitude) $ x_i $ and auxiliary variable (error pulse amplitude) $ e_i $ obey the following deterministic equations:{\scriptsize $^{\cite{Leleu2019}}$}

\begin{equation}{\label{eq:i-th spin 1}}
	\frac{dx_i}{dt} = -x_i^3 + \left( p - 1 \right) x_i - e_i \sum_j \xi  J_{ij} x_j,
\end{equation}

\begin{equation}{\label{eq:i-th spin 2}}
	\frac{de_i}{dt} = -\beta e_i \left( x_i^2 - \alpha\right),
\end{equation}

where $ J_{ij} $ is the Ising coupling matrix, $\alpha$, $\beta$ and $p$ are system parameters and $\xi$ is a normalizing constant for $J_{ij}$ (see Appendix A for parameter selection).
In many cases, we may modulate these parameters over time to achieve better performance (see Section 3 and Appendix C). 
To use this system as an Ising solver we consider the spin configuration $\sigma_i = \textrm{sign}(x_i)$ as a possible solution to the corresponding Ising problem. 
Even though noise is ignored in the above-mentioned equation, we can choose the initial $x_i$ amplitude randomly to create a diverse set of possible trajectories.

\medskip
In this paper, we refer to this system of equations as CIM with chaotic amplitude control (CIM-CAC). The term ``chaotic” is used because CIM-CAC exhibits chaotic behavior (as discussed in Section 3). CIM-CAC may refer to either the above-mentioned system of deterministic differential equations when integrated as a digital algorithm or an optical CIM that emulates the above-mentioned equations.

\medskip
While studying the CIM-CAC equations, we have made the following modification:
\begin{equation}{\label{eq:CIM-CAC 1}}
	z_i =  e_i \sum_j \xi  J_{ij}x_j,
\end{equation}
\begin{equation}{\label{eq:CIM-CAC 2}}
	\frac{dx_i}{dt} = -x_i^3+\left( p-1\right) x_i - z_i,
\end{equation}
\begin{equation}{\label{eq:CIM-CAC 3}}
	\frac{de_i}{dt} = -\beta e_i\left( z_i^2 - \alpha\right),
\end{equation}

which we refer to as CIM with chaotic feedback control (CIM-CFC). The only difference between Eqs. (\ref{eq:i-th spin 2}) and (\ref{eq:CIM-CAC 3}) is that the time evolution of the error variable $e_i$ monitors the feedback signal $z_i$, rather than the internal amplitude $x_i$. 
The dynamics of this new equation are very similar to those of CIM-CAC, which can be understood by observing that CIM-CAC and CIM-CFC have nearly identical fixed points. 
The motivation for studying CIM-CFC in addition to CIM-CAC is to gain a better understanding of how these systems work. 
In addition, CIM-CFC may have slightly simpler dynamics, which simplifies its numerical integration.

\medskip
The third system discussed in this paper has a very different equation:
\begin{equation}{\label{eq:third system 1}}
	z_i =  \sum_j \xi  J_{ij} x_j,
\end{equation}
\begin{equation}{\label{eq:third system 2}}
	\frac{dx_i}{dt} = -x_i^3 + \left( p-1 \right) x_i - f \left( cz_i \right) - k \left( z_i - e_i \right),
\end{equation}
\begin{equation}{\label{eq:third system 3}}
	\frac{de_i}{dt} = -\beta \left( e_i - z_i \right).
\end{equation}
The non-linear function $f$ is a sigmoid-like function such as $f(z) = \tanh(z)$, and $p$, $k$, $c$ and $\beta$ are system parameters (See Appendix A for parameter selection). The significance of this new feedback system is that the differential equation for the error signal $e_i$ is now linear in the ``mutual coupling signal" $z_i$. 
In addition, $z_i$ is calculated simply as $\sum_j \xi J_{ij} x_j$ without the additional factor $e_i$ as in Eq. (\ref{eq:third system 1}). 
This means that the only nonlinear elements in this system are the gain saturation term $ -x_i^3 $ and the nonlinear function $f$. 
For the results in this paper we will use $f(z) = \tanh(z)$, however if a different function with the same properties is used the system will have similar behavior.

\medskip
In the above-mentioned  system, the two essential aspects of CIM-CAC and CIM-CFC are separated into two different terms.
The term $f \left( cz_i \right)$ realizes mutual coupling while passively addressing the problem of amplitude heterogeneity, while the term $k \left( z_i - e_i \right)$ introduces the error signal $e_i$ which helps to destabilize local minima. 
Therefore, we refer to this system as CIM with separated feedback control (CIM-SFC) in the remainder of this paper.

\medskip
Another significant aspect of CIM-SFC (Eqs. (\ref{eq:third system 1}),(\ref{eq:third system 2}) and (\ref{eq:third system 3})) compared to CIM-CAC and CIM-CFC (Eqs. (\ref{eq:i-th spin 1})-(\ref{eq:CIM-CAC 3})) 
is that the auxiliary variables $e_i$ in CIM-SFC have a very different meaning. 
In CIM-CAC and CIM-CFC, $e_i$ is meant to be a strictly positive number that varies exponentially and modulates the mutual coupling term. 
In CIM-SFC, $e_i$ is instead a variable that stores sign information and takes the same range of values as the mutual coupling signal $z_i$. 
The error signal $e_i$ can essentially be regarded as a low pass filter on $z_i$, and the term $k (e_i - z_i)$ can be regarded as a high pass filter on $z_i$ (in other words $k (e_i - z_i)$ only registers sharp changes in $z_i$). 
The similarities and differences among CIM-SFC, CIM-CAC and CIM-CFC can be understood by observing the fixed points. In CIM-CAC and CIM-CFC, the fixed points are of the form:{\scriptsize $^{\cite{Leleu2019}}$}
\begin{equation}{\label{eq:CIM-CAC and CFC 1}}
	x_i = \lambda_1 \sigma_i,
\end{equation}
\begin{equation}{\label{eq:CIM-CAC and CFC 2}}
	e_i = \lambda_2 \frac{1}{h_i \sigma_i},
\end{equation}
with
\begin{equation}{\label{eq:CIM-CAC and CFC 3}}
	h_i = \sum_j \xi  J_{ij} \sigma_j.
\end{equation}
Here, $\sigma_i$ is a spin configuration corresponding to a local minimum of the Ising Hamiltonian, and $\lambda_1$ and $\lambda_2$ are constants that depend on the system parameters. 
In CIM-SFC, the fixed points are generally very complicated and difficult to express explicitly. 
However, if we consider the limit of $ c \gg 1 $, the fixed points will take the form:
\begin{equation}{\label{eq:c >> 1}}
	x_i = \lambda\sigma_i,
\end{equation}
\begin{equation}{\label{eq:c >> 2}}
	e_i = \lambda h_i,
\end{equation}
where $\lambda$ is a number such that $-\lambda^3  + (p-1)\lambda = -1$.
Again, $\sigma_i$ is a spin configuration corresponding to a local minimum. This formula is only valid if $f(cz)$ is an odd function that takes the value of $+1$ for $ cz \gg 1 $ and $-1$ for $ cz \ll -1 $. 
Therefore, it is important to choose an appropriate function $f$.

\medskip
The important difference between the fixed points of these two types of systems is that in CIM-CAC and CIM-CFC, the error signal is
\begin{eqnarray}
	\left| e_i \right|  \propto \frac{1}{ \left| h_i \right| }, \nonumber
\end{eqnarray}
whereas in CIM-SFC, the error signal is
\begin{eqnarray}
\left| e_i \right|  \propto \left| h_i \right|. \nonumber
\end{eqnarray}
In Section 5, we will see that this difference makes CIM-SFC more robust to quantum noise from reservoirs and pump sources. 
In the next section, we will investigate the similarities and differences among these three systems using numerical simulation.

\section{Numerical Simulation of CIM-CAC, CIM-CFC and CIM-SFC}

The originally proposed CIM architecture employs simple linear feedback without using an error detection/correction mechanism. In other words, the feedback term in Eq. (\ref{eq:i-th spin 1}) is simply $\sum_{j}{  \xi  J_{ij} x_j} $.{\scriptsize $^{\cite{Wang2013}}$} 
In this case, the Ising Hamiltonian cannot be properly mapped to the network loss owing to OPO amplitude heterogeneity, especially for frustrated spin systems, as shown in Figure \ref{fig:three CIMs} (a). 
Such a CIM often does not find a ground state of the Ising Hamiltonian; instead, it selects the lowest-energy eigenstate of the coupling (Jacobian) matrix [$ J_{ij} $].\cite{Wang2013} 
This undesired operation is caused by a system's formation of heterogenous
amplitudes.\cite{Wang2013}  We can address this problem partially by introducing a nonlinear filter function for the feedback pulse, such as 
$ \tanh ( \sum_{j}\xi J_{ij}x_j) $. Thus, we can achieve the homogeneous OPO amplitudes, at least well above threshold,  as shown in Figure \ref{fig:three CIMs} (b), and satisfy the proper mapping condition toward the end of the system’s trajectory. 
However, such nonlinear filtering alone is not sufficiently powerful to prevent the machine from being trapped in numerous local minima. 
As the problem size N increases, NP-hard Ising problems are expected to have an exponentially increasing number of local minima; hence, a system that is easily trapped in these minima will be ineffective.

\begin{figure}[!htb]
	\centering
	\includegraphics[scale=0.1]{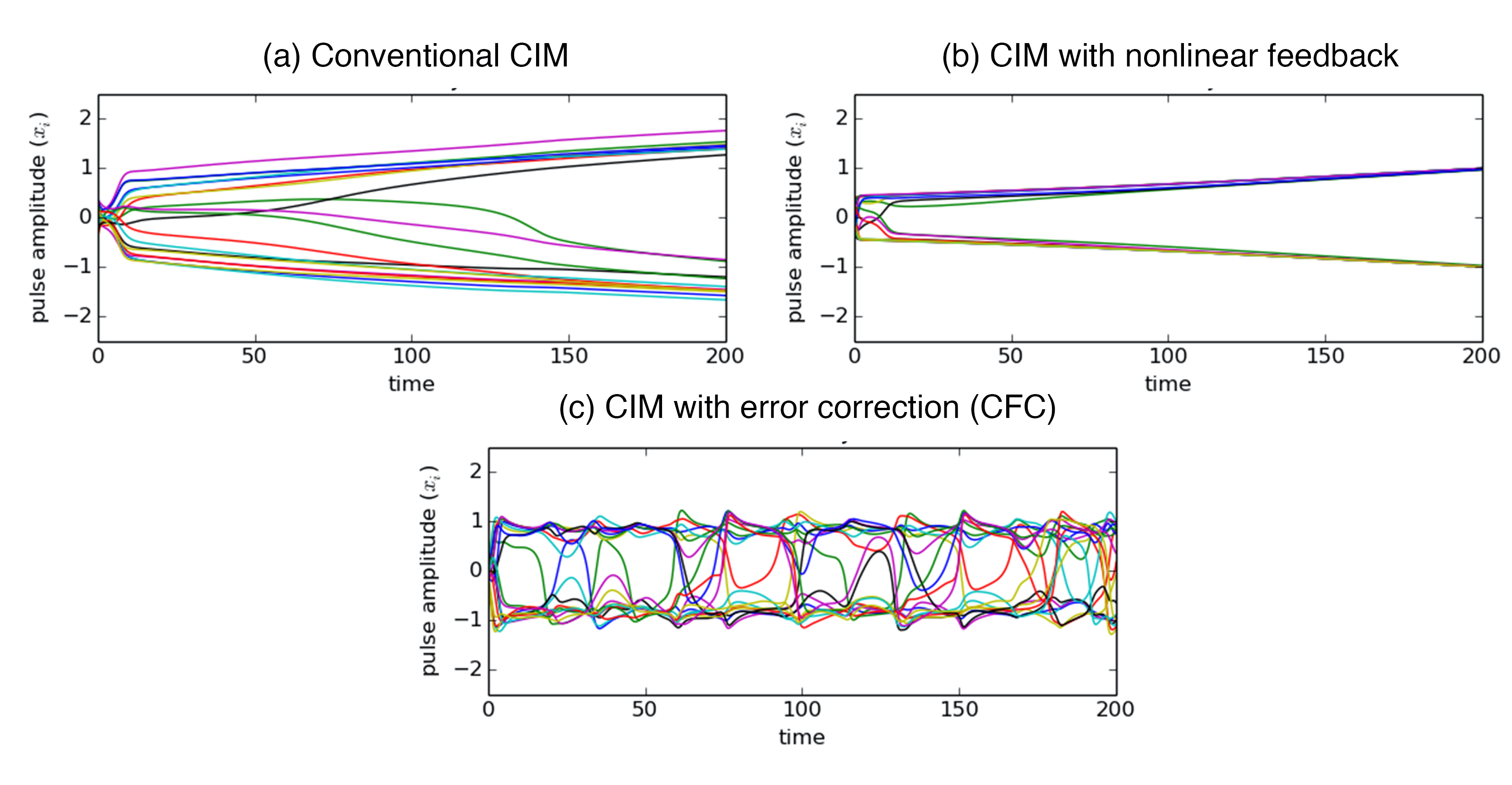}
	\caption{Trajectories of OPO amplitudes in CIMs with (a) linear feedback, (b) nonlinear filtering feedback, and (c) chaotic feedback control.}
	\label{fig:three CIMs}
\end{figure}

\medskip
To destabilize the attractors caused by local minima and allow the machine to continue searching for a true ground state, we can introduce an error detection/correction variable expressed by Eq. (\ref{eq:i-th spin 2}) or (\ref{eq:CIM-CAC 3}).
As shown in Figure \ref{fig:three CIMs} (c), the trajectory of a CIM with error correction variables will not reach equilibrium but continue to explore many states. 
Conversely, the systems in Figure \ref{fig:three CIMs} (a) and (b), which do not have an error correction variable ($e_i$), will often converge rapidly on a fixed point corresponding to a high-energy excited state of the Ising Hamiltonian.
Destabilizing the local minima will inevitably make the ground state unstable as well. 
Although this is undesirable, we can simply allow the machine to visit many local minima and then determine the one that has the lowest energy subsequently. 
Alternatively, we have found that by modulating the system parameters, the system will have a high probability of staying in a ground state toward the end of the trajectory (see Section 4 for further details).

\medskip

The addition of another N degrees of freedom allows the machine to visit a local minimum, escape from it, and continue to explore nearby states; this is not possible with a conventional CIM algorithm. 
In this section, we will discuss the dynamics of the error correction schemes proposed in this paper: CIM-CFC and CIM-SFC.



\medskip
Even though CIM-CFC and CIM-SFC are described by very different equations, the two systems were originally conceived through a similar concept.
To understand why CIM-CFC and CIM-SFC are similar, we can consider these systems as follows. We introduce the ``mutual coupling signal" $ M_i(t) = \sum_j \xi J_{ij} x_j (t) $ and the ``injection feedback signal" $ I_i (t) $. Then, we can write both CIM-CFC and CIM-SFC in the form:
\begin{equation}
	M_i(t) = \sum_j \xi J_{ij} x_j(t),
\end{equation} 
\begin{equation}
	\frac{dx_i}{dt} = -x_i^3 + \left( p - 1 \right) x_i - I_i (t),
\end{equation}
where $ I_i (t)$ depends on the time evolution of $ M_i (t) $. Figure \ref{fig:feedback systems} shows how $ I_i (t) $ (red) varies with respect to a mutual coupling field $ M_i (t) $ (blue) for four different feedback schemes.

\begin{figure}[!htb]
	\centering
	\includegraphics[scale=0.1]{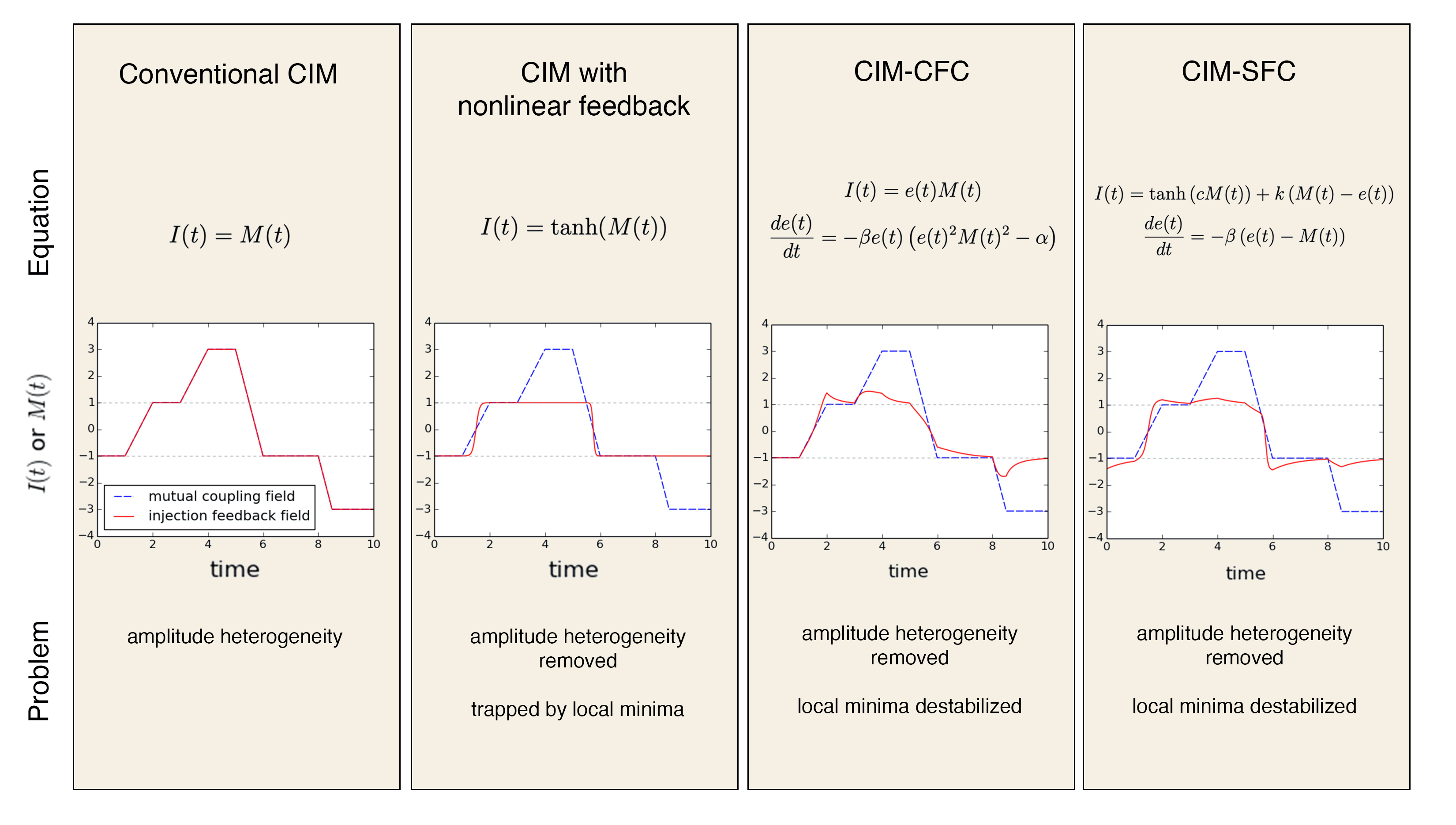}
	\caption{Mutual coupling field (blue) and injection feedback field (red) in four different feedback systems.}
	\label{fig:feedback systems}
\end{figure}

The similarity between CIM-CFC and CIM-SFC is as follows: if the mutual coupling field  $ M_i (t) $ remains constant for a certain period of time, then the injection feedback field $ I_i (t) $ will converge on the value given by $\textrm{sign}(M_i (t))$. 
However, if $M_i (t)$ varies sharply, then $I_i (t)$ will deviate from its steady state values: $ +1/-1 $. 
This small deviation is effective for triggering destabilization when the system is near a local minimum, which allows the machine to explore new spin configurations.

\medskip
Although CIM-CFC and CIM-SFC were conceived on the basis of the same principle, the dynamics of the two systems seem to differ from each other. 
In particular, CIM-CFC (and CIM-CAC) nearly always features chaotic dynamics, as the trajectory is highly sensitive to the initial conditions. 
In the case of CIM-SFC, the trajectory will often immediately fall into a stable periodic orbit unless the parameters are dynamically modulated. 
At present, we do not have an exact theoretical reason for this difference in dynamics; this is purely an experimental observation. 
A more theoretical analysis in the case of CIM-CAC can be found elsewhere. {\cite{Leleu2019}}.

\medskip
To demonstrate this difference, Figure \ref{fig:correlations} shows the correlation of pulse amplitudes between two initial conditions that are very close to each other. 
An initial condition for the pulse amplitude \#1 (plotted on the x-axis) is chosen from a zero-mean Gaussian with a standard deviation of 0.25, while the other initial condition for trajectory \#2 (plotted in y-axis) is equal to that of trajectory \#1 plus a small amount of noise (standard deviation 0.01).

\begin{figure}[!htb]
	\centering
	\includegraphics[scale=0.1]{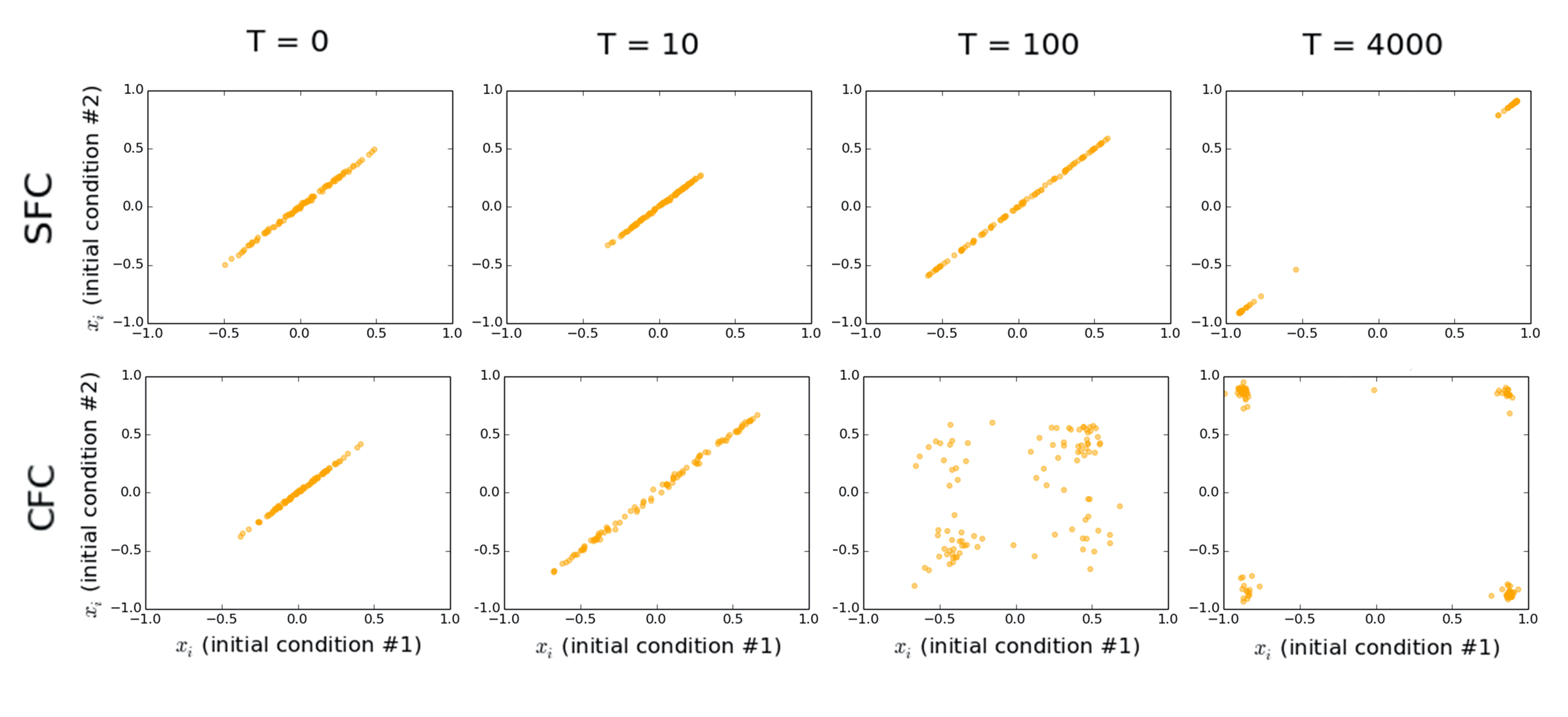}
	\caption{Signal pulse amplitude correlations at different evolution time in CIM-SFC and CIM-CFC.}
	\label{fig:correlations}
\end{figure}

\medskip
Figure \ref{fig:correlations}, shows the correlation of all 100 pulse amplitudes between the two initial conditions for a Sherrington-Kirkpatrick (SK) spin glass instance of problem size $N = 100$. 
In CIM-SFC (first row), the correlation remains even after 4000 time steps (round trips), which means that two initial conditions follow a nearly identical trajectory. 
However, in CIM-CFC (second row), we see that the xi variables become uncorrelated after around 100 time steps, even though the initial conditions of the two trajectories are very close. 
This indicates qualitatively that CIM-CFC is highly sensitive to the initial condition, whereas  CIM-SFC is not.

\medskip
This pattern tends to hold when different parameters and initial conditions are used. 
However, although CIM-SFC stays correlated in most cases, the two trajectories diverge under certain system parameters and initial conditions. 
This means that although CIM-SFC is less sensitive to the initial conditions compared to CIM-CFC, some chaotic dynamics likely occur during the search, especially when the parameters are modulated.

\medskip
Another way to qualitatively observe the difference in dynamics is to simply observe the trajectories.
Figure \ref{fig:trajectories}, shows examples of trajectories of both systems (10 out of 100 $x_i$ variables are shown) with fixed and linearly modulated system parameters.
When the parameters are fixed, the difference between the two systems evident. 
CIM-SFC will rapidly become trapped in a stable periodic attractor, while CIM-CFC will continue to search in an unpredictable manner. 
Therefore, the parameters are slowly modulated in CIM-SFC so that the system can find a ground state. 
CIM-CFC and CIM-CAC can find ground states with fixed parameters. 
However, we have found that modulation of the system parameters improves the performance of CIM-CFC and CIM-CAC considerably (see Appendix C for details).

\begin{figure}[!htb]
	\centering
	\includegraphics[scale=0.1]{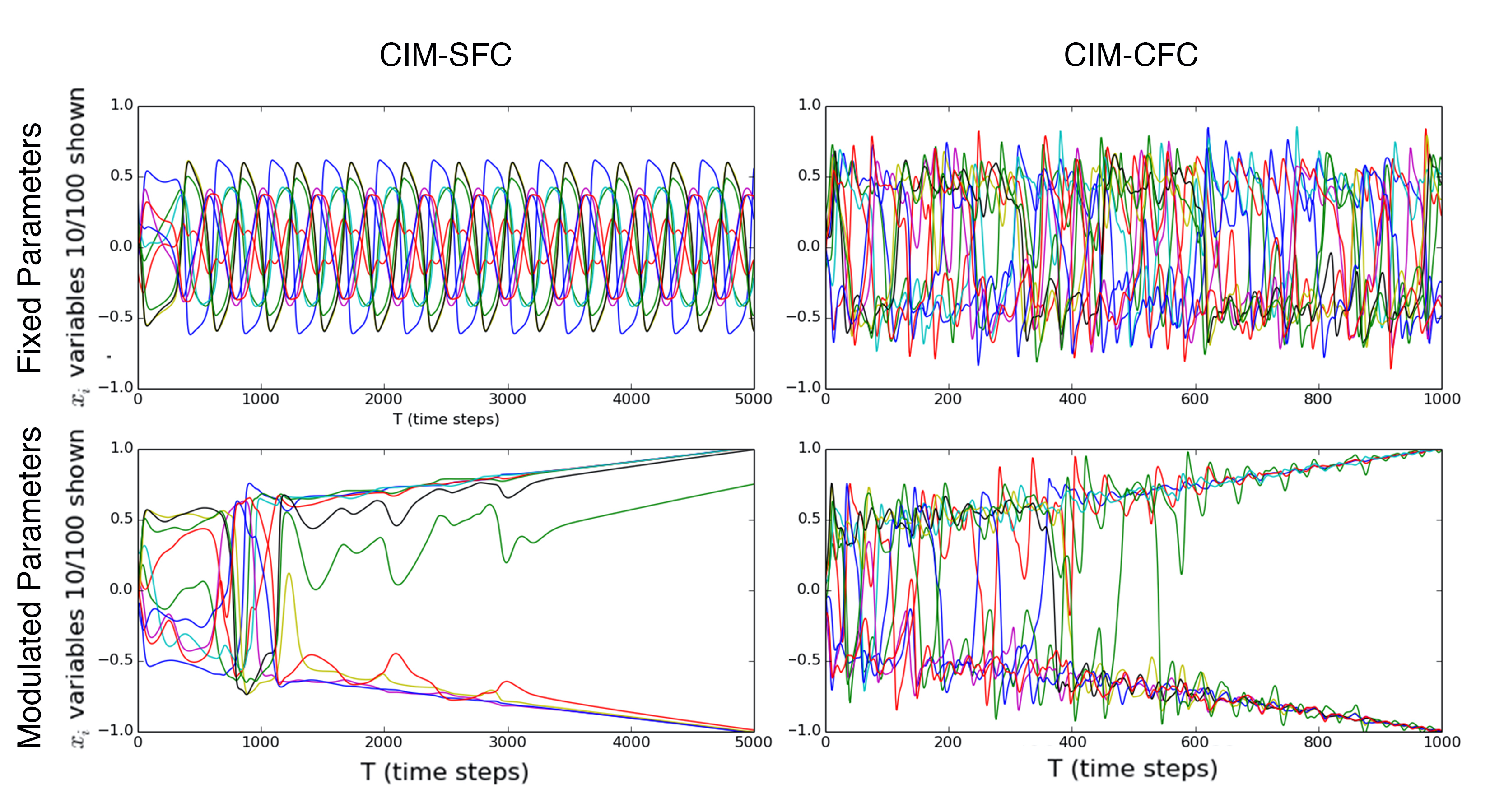}
	\caption{Signal pulse amplitude trajectories of CIM-SFC and CIM-CFC with fixed and modulated system parameters.}
	\label{fig:trajectories}
\end{figure}

\medskip
In the lower left panel of Figure \ref{fig:trajectories}, the parameters $c$ and $p$ of CIM-SFC are linearly increased from low to high values ($p$ ranges from $ -1 $ to $ +1 $ and $c$ ranges from 1 to 3). We can see that as the parameters change, the system may jump from one attractor to another and eventually end up in a fixed point/local minimum. 
By linearly increasing the parameters $c$ and $p$ from a low to high value in CIM-SFC, we are slowly transitioning the nonlinear term $\tanh(cz_i)$ from a ``soft spin" mode where the nonlinear coupling term has a continuous range of values between $ -1 $ and $1$ to a ``discrete" mode where $\tanh(cz_i)$ will mostly take on the values $ +1 $ or $ -1 $. 
This transition is seems to be crucial for CIM-SFC to function properly.

\medskip
For most fixed parameters, CIM-SFC rapidly approaches a periodic or fixed point attractor as shown in Figure \ref{fig:trajectories}; however, as mentioned earlier it is likely that for some specific values of $c$ and $p$, CIM-SFC 
will feature chaotic dynamics similar to CIM-CFC. It has been shown{\scriptsize $^{\cite{Leleu2019, Ercsey-Ravasz2011}}$} that chaotic dynamics are observed when solving hard optimization problems efficiently using a deterministic system.
This trend is also observed in the simulated bifurcation machine \cite{Goto2019, Goto2021}. Whether or not CIM-SFC utilizes chaotic dynamics is beyond the scope of this paper.
Whether CIM-SFC uses chaotic dynamics is beyond the scope of this paper. To answer this question, we need to further analyze how the parameters affect the dynamics of CIM-SFC and gain a deeper understanding of how CIM-SFC finds ground states.

\section{Implementation of CIM with Optical Error Correction Circuits}
Figure \ref{fig:all-optical CIM}, together with Figure \ref{fig:semi-classical}(c), shows a physical setup for CIM-CAC and CIM-CFC with optical error correction circuits. 
In our design, the main ring cavity stores both signal pulses with normalized amplitude, $x_i$ and error pulses with normalized amplitude $e_i$ , where $i=1,2, \cdots N$. 
The signal pulses start from vacuum states $\left| 0 \right\rangle _1 \left| 0 \right\rangle _2 \cdots \left| 0 \right\rangle _N$ and are amplified (or deamplified) along the X-coordinate by a positive (or negative) pump rate $p$. 

\begin{figure}[!htb]
	\centering
	\includegraphics[scale=0.5]{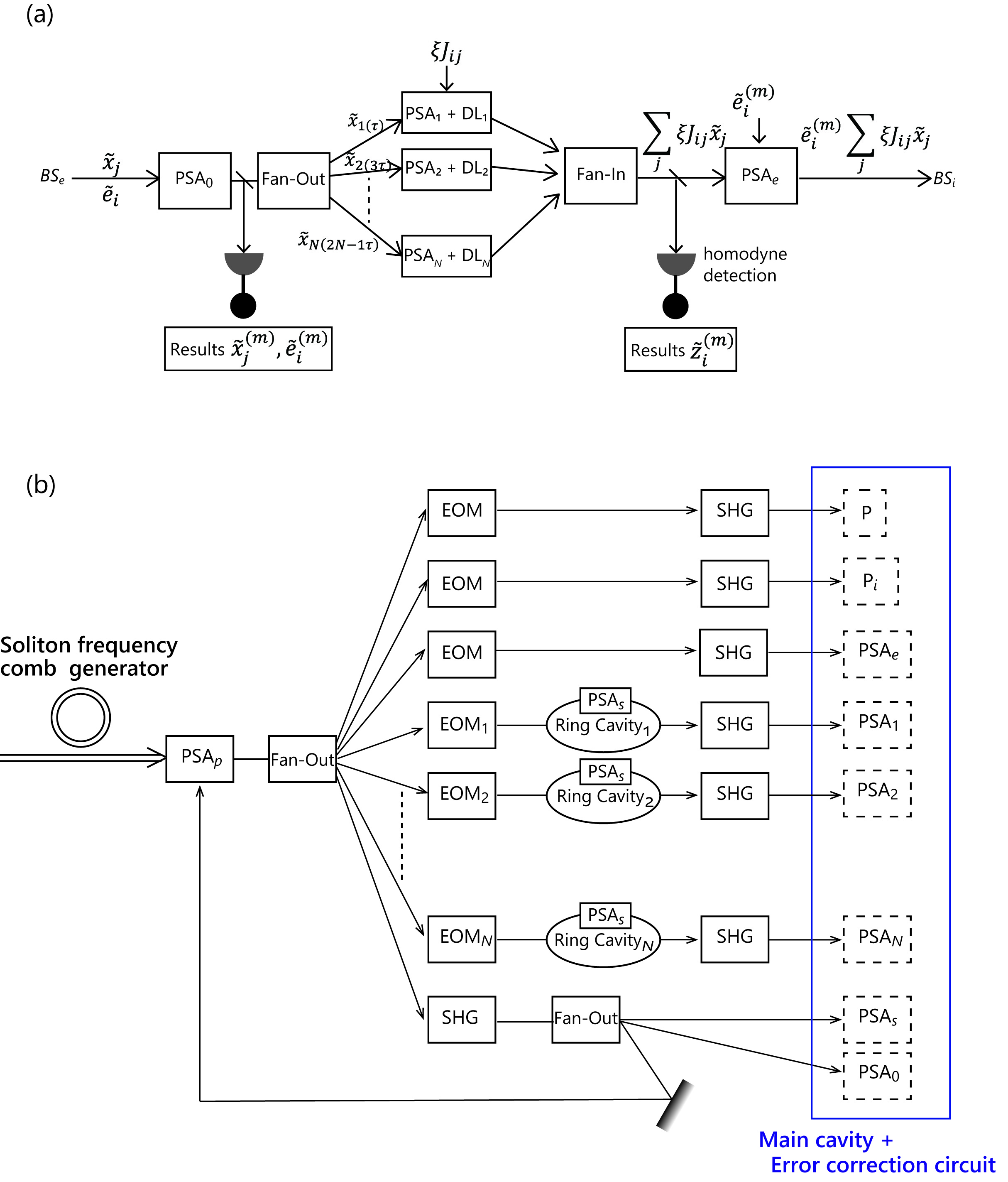}
	\caption{(a) Optical implementation of error correction circuits for CIM-CAC and CIM-CFC. (b) Pump pulse factory providing SHG pulses to the main cavity, and the error correction circuits. 
	The pump pulse factory carries $N^2$ pulses spread over $N$ optical cavities corresponding to the elements of the Ising coupling matrix $J_{ij}$.}
	\label{fig:all-optical CIM}
\end{figure}

\medskip
The error pulses start from a coherent state $\left| \alpha \right\rangle _1 \left| \alpha \right\rangle _2 \cdots \left| \alpha \right\rangle _N$ with $\alpha > 0$ and are amplified
 (or deamplified) along the X-coordinate by the pump rate $p'$ as described below. 
 The squared amplitude of the error pulses is kept small ($e_i^2 < 1$) compared to the saturation level of the main cavity OPO. Thus the 
 error pulses are controlled in a linear amplifier/deamplifier regime while the signal pulses are controlled in both a linear amplifier/deamplifier regime ($x_i^2 < 1$) and a nonlinear oscillator regime ($x_i^2 > 1$).

\medskip
An extraction beamsplitter (BS$_e$ shown in Figure \ref{fig:semi-classical}(c)) selects partial waves of the signal and error pulses that are amplified by a noise-free phase-sensitive amplifier (PSA$_0$ as shown in Figure \ref{fig:all-optical CIM}(a)). 
PSA$_0$ amplifies the signal and error pulses to a classical level without introducing additional noise.
The extracted amplitudes $\tilde{x_i}$ and $\tilde{e_i}$ suffer from the signal-to-noise ratio (SNR) degradation owing to the vacuum noise incident on BS$_e$. 
However, they are amplified by a high-gain noise-free phase-sensitive amplifier PSA$_0$ to classical levels; hence, no further SNR degradation occurs even with large linear losses in the optical error correction circuits.

\medskip
A small part of the PSA$_0$ output is sent to an optical homodyne detector that measures the extracted signal and error pulses with amplitudes $\tilde{x_i}$ and $\tilde{e_i}$, respectively.
The measurement error of the homodyne detection is determined solely by the reflectivity of BS$_e$ and the vacuum fluctuation incident on BS$_e$ (as described above).
Figure \ref{fig:all-optical CIM}(a) shows the output of the fan-out circuit at different time instances $t = {\tau, 3\tau, 5\tau, ...}$ separated by a signal pulse to signal pulse interval of $2\tau$.

\medskip
For instance, the signal pulse ($\tilde{x_j}$) is first input into PSA$_j$ and then sent to optical delay line DL$_j$ with a delay time of $(2N - 2j + 1)\tau$.
The phase-sensitive gain/loss of PSA$_j$ is set to $\sqrt{G_j} = \xi J_{ij}$ so that the amplified/deamplified signal pulse that arrives in front of the fan-in circuit at time 
$t = 2N\tau$ is equal to $\xi J_{ij} \tilde{x_j}$. Therefore the fan-in circuit will output a pulse with the desired amplitude of $\sum_j \xi J_{ij} \tilde{x_j}$.
Suppose that PSA$_j$ has a phase-sensitive linear gain/loss of 10dB, then we can implement an arbitrary Ising coupling of range $10^{-2} < |\xi J_{ij}| < 1$.

\medskip
Next, the output of the fan-in circuit is input into another phase-sensitive amplifier PSA$_e$
that amplifies with a factor of $\sqrt{G_e} = \tilde{e_i}$. 
This is achieved by modulating the pump power to PSA$_e$ based on the measurement result for $\tilde{e_i}$.
Finally, the output of PSA$_e$ is injected back into signal pulse ($x_i$) of the main cavity via BS$_i$ (see Figure \ref{fig:semi-classical}(c)).
The extraction beamsplitter BS$_e$ outputs not only signal pulses but also error pulses that are used only for homodyne detection. 
Thus, we switch off the pump power to PSA$_0$ for the error pulses and deamplify the residual error pulses by PSA$_1$, PSA$_2$, ... , PSA$_N$, PSA$_e$.
In this way we avoid any spurious injection of the error pulse back into the main cavity. The dynamics of the error pulse are governed solely by the pump power $p_i$ to the main cavity PSA, which is set to 
satisfy $p_i - 1 = \beta \left(\alpha - \tilde{x_i}^2\right)$ or $p_i - 1  = \beta \left(\alpha - \tilde{z_i}^2\right)$.

\medskip
One advantage of this optical implementation of CIM-CAC and CIM-CFC is that only one type of active device, a noise-free phase-sensitive (degenerate optical parametric) amplifier, 
and all the other elements are passive devices. 
This fact may allow for on-chip monolithic integration of the CIM system as well as low-energy dissipation in the computational unit, which will be discussed in Section 5.

\medskip
A similar optical implementation of CIM-SFC is shown in Appendix F.

\medskip
Figure \ref{fig:all-optical CIM}(b) shows a pump pulse factory that provides the second harmonic generation (SHG) pulses to the main cavity PSA, post-amplifier PSA$_0$, delay line amplifiers PSA$_1$, PSA$_2$, $\cdots$ PSA$_N$ and exit amplifier PSA$_e$. 
The purpose of this pump pulse factory is to reduce the use of EOM modulators, which consume the most energy in the entire CIM. 
A soliton frequency comb generator produces a pulse train at a repetition frequency of 100 GHz and wavelength of 1.56 $\mu\textrm{m} $ wavelength. 
Before it is split into many branches, the pulse train is amplified by a pump amplifier PSA$_p$. $N$ storage ring cavities continuously produce the pump pulses for PSA$_1$, PSA$_2$, $\cdot$ PSA$_N$ in order to implement the MVM $ \sum \xi J_{ij} \tilde{x}_j $.
For this purpose, the pulses stored in the i-th ring cavity acquire the appropriate amplitudes to realize the gain $ \sqrt{G_{ij}} = \xi J_{ij} $.
The time duration for using $N$ EOM arrays is only one round trip of the ring cavity, i.e., $ N \times 10 $ (psec).
The out-coupling loss of the storage ring cavities is compensated for by the linear gain of the internal PSA$_s$. The pump pulses for PSA$_p$, PSA$_s$, and PSA$_0$have constant amplitudes and are hence driven directly by the PSA$_p$ output. 
The pump pulses $ P $ and $ P_i $ for the signal and error pulses in the main cavity, as well as the exit PSA$_e$, must be modulated during the entire computation time.

\medskip
Another detail that needs to be accounted for when considering an optical implementation is the calculation of the Ising energy. 
In our digital simulation for generating the results presented in this paper, the Ising energy is calculated at every time step (round trip), and the smallest energy obtained is used as the result of the computation.
This means that, in an optical implementation, we must measure the $\tilde{x_i}$ amplitude in every round trip and calculate the Ising energy using, for instance, an external ADC/FPGA circuit. 
This would defeat the purpose of using optics, as the digital circuit in the ADC/FPGA would then become a bottleneck in terms of time and energy consumption.

\medskip
However, we have found that with proper parameter modulation as shown in Figure \ref{fig:trajectories}, it is possible to use only the final state of the system for the result and still have a high success probability.
For the results on 800-spin Ising instances (SK model) presented in Section 6, we calculated how often a successful trajectory is in the ground spin configuration after the final time step.  
We found that, for CIM-SFC, in 100\% of the 7401 successful trajectories the final spin configuration was in the ground state. 
In other words, if CIM-SFC visits the ground state at any point during the trajectory, then it will also be in the ground state at the end of the trajectory.
Meanwhile, for CIM-CFC and CIM-CAC, this was true  only in 75\% of the time and 48\% of the time, respectively. 
We believe that this difference among the three systems is a result of both the intrinsic dynamics and the parameters used.

\medskip 
This suggests that in a CIM with optical error correction, we can simply digitize the final measurement result of $x_i$ after many round trips to obtain the computational result, and still have a high success probability. 
In the case of CIM-CFC and CIM-CAC, it might be beneficial to read the spin configuration multiple times during the last few round trips, as the machine usually visits the ground state close to the end of the trajectory even if it does not stay there.

\section{Quantum Noise Analysis and Energy Cost to Solution}

\label{quantum noise}
As we propose implementation of these dynamical systems on analog optical devices, it is important to investigate the extent to which the noise from the physical systems (in this case, quantum noise from pump sources and external reservoirs) will degrade the performance. 
In this section, we present quantum models based on our optical implementation.

\medskip
In our optical implementation for CIM-CAC, the real-number signal pulse amplitude $ \mu_i$ (in unit of photon amplitude) (in units of photon amplitude) obeys the following truncated Wigner SDE:{\scriptsize $^{\cite{Maruo2016, Inui2020}}$} 
\begin{equation}{\label{eq:Wigner SDE}}
	\frac{d}{dt} \mu_i = \left( p - 1 \right) \mu_i - g^2 \mu^3_i + \tilde{\nu}_i  \sum_{j}{\xi J_{ij}} \tilde{\mu}_j + n_i,
\end{equation}
where the term $p \mu_i$ represents the parametric linear gain and the term $ - \mu_i $ represents the linear loss rate; this includes the cavity background loss and extraction/injection beam splitter loss for mutual coupling and error correction. 
The nonlinear term $ - g^2 \mu^3_i $ represents gain saturation (or back-conversion from signal to pump), where $g$ is the saturation parameter. 
The saturation photon number is given by $ 1/g^2 $, which is equal to the average photon number of a solitary OPO at a pump rate of $ p=2 $ (two times above the threshold). 
Furthermore, $ J_{ij} $ is the ($ i, j $) element of the $ N \times N $ Ising coupling matrix, as described in Section 2. 
The time $t$ is normalized by a linear loss rate; hence, the signal amplitude decays by a factor of $1/e$ at time $t=1$.
In addition, $ \tilde{\mu}_j = \mu_j + \Delta \mu_j $ and $ \tilde{\nu}_i = \nu_i + \Delta \nu_i $ are the inferred amplitudes for the signal pulse and error pulse, respectively, and $ \Delta \mu_j $ and $ \Delta \nu_i $ represent the additional noise governed by vacuum fluctuations incident on the extraction beam splitter. 
They are characterized by $ \sqrt{\frac{ 1 - R_B}{4 R_B}} w $, where $ R_B $ is the reflectivity of the extraction beam splitter and  $ w $ is a zero-mean Gaussian random variable with a variance of one. Finally, $ n_i $ is the noise injected from external reservoirs and pump sources.{\scriptsize $^{\cite{Maruo2016, Inui2020}}$}  
It is characterized by the two time correlation functions $ \left\langle n_i (t) n_i (t^{\prime}) \right\rangle  = (\frac{1}{2} + g^2 \mu^2_i) \delta (t - t^{\prime})$.
We assume that the external reservoirs are in vacuum states and that the pump fields are in coherent states.

\medskip
The real number error pulse amplitude $ \nu_i $ (in units of photon amplitude) is governed by
\begin{equation}{\label{eq:error detection}}
	\frac{d}{dt} \nu_i = \left( p^{\prime}_{i} - 1 \right) \nu_i + m_i,
\end{equation}
where the correlation function for the noise term is given by $ \left\langle m_{i} (t) m_{i} (t^{\prime}) \right\rangle  = \frac{1}{2} \delta (t - t^{\prime}) $. 
The pump rate $ p^{\prime}_i $ for the error pulse is determined by the inferred signal pulse amplitude $ \tilde{x}_i = g \tilde{\mu}_i $ normalized by the saturation parameter,
\begin{equation}{\label{eq:normalized signal pulses}}
	p^{\prime}_{i}-1 = \beta \left( \alpha - \tilde{x}^2_i \right).
\end{equation}

\medskip
The error pulses start from coherent states $\left| \gamma \right\rangle _1 \left| \gamma \right\rangle _2 \cdots \left| \gamma \right\rangle _N$, for some positive real number $1/g \gg \gamma > 0$. 
The absence of a gain saturation term in Eq. (\ref{eq:error detection}) implies that the error pulses are always pumped at below the threshold.
Nevertheless, the error pulses represent exponentially varying amplitudes.

\medskip
The parameter $ \beta $ governs the time constant for the error correction dynamics, and $ \alpha $ is the squared target amplitude. This feedback model stabilizes the squared signal pulse amplitude $ \tilde{x}^2_i = g^2 \tilde{\mu}^2_i $ to $ \alpha $ through an exponentially varying error pulse amplitude $ e_i = g \nu_i $. 
Eqs. (\ref{eq:Wigner SDE}) and (\ref{eq:error detection}) are rewritten for the normalized amplitudes $ x_i $ and $ e_i $ as

\begin{equation}{\label{eq:Wigner SDE 1}}
	\frac{d}{dt} x_i = \left( p - 1 \right) x_i - x^3_i  + \tilde{e}_i \sum_{j}{ \xi J_{ij}} \tilde{x}_j + gn_i,
\end{equation}
\begin{equation}{\label{eq:error detection 1}}
	\frac{d}{dt} e_i = \left( p^{\prime}_i - 1 \right) e_i + gm_i.
\end{equation}
which are nearly identical to Eqs. (\ref{eq:i-th spin 1}) and (\ref{eq:i-th spin 2}) except for the noise terms.

\medskip
CIM-CFC is also realized using the experimental setup shown in Figure \ref{fig:all-optical CIM}. In this case, the relevant truncated
Wigner SDE for the error pulse amplitude is still given by Eq. (\ref{eq:error detection}) or (\ref{eq:error detection 1}); however, the pump rate $ p^{\prime} $ should be modified to
\begin{equation}{\label{eq:effective pump rate p'}}
	p^{\prime}_{i}-1 = \beta \left( \alpha - \tilde{z}^2_i \right).
\end{equation}
with
\begin{equation}{\label{eq:equation for z_i}}
	\tilde{z}_i = \sum_{j}{ \xi J_{ij}} \tilde{x}_j
\end{equation}

\medskip
Finally, CIM-SFC can also be realized using the experimental setup shown in Appendix F (Figure \ref{fig:all-optical CIM 2}). In this case, Eqs. (\ref{eq:Wigner SDE 1}) and (\ref{eq:error detection 1}) should be modified as
\begin{equation}{\label{eq:Zi SDE 2}}
	\tilde{z}_i = \sum_{j}{ \xi J_{ij}} \tilde{x}_j
\end{equation}
\begin{equation}{\label{eq:Wigner SDE 2}}
	\frac{d}{dt} x_i = \left( p - 1 \right) x_i - x^3_i  + k \left( \tilde{e}_i -  \tilde{z}_i \right) + \tanh \left( c \tilde{z}_i \right) + gn_i,
\end{equation}
\begin{equation}{\label{eq:error detection 2}}
	\frac{d}{dt} e_i = - \beta \left( e_i -  \tilde{z}_i \right) + gm_i.
\end{equation}

If we compare the semi-classical nonlinear dynamical models of CIM-CAC, CIM-CFC, and CIM-SFC, represented by Eqs. 
(\ref{eq:i-th spin 1})-(\ref{eq:third system 3}), with the quantum nonlinear dynamical models (truncated Wigner SDE), represented by Eqs. (\ref{eq:Wigner SDE 1})-(\ref{eq:error detection 2}),
 we find that the main difference is the absence or presence of the vacuum noise and pump noise terms $ gn_i $ and $ gm_i $, respectively. 
 The other important difference is that $\tilde{x_i}$ and $\tilde{e_i}$ are inferred amplitudes with the vacuum noise contribution in the quantum model, whereas in the semi-classical model, the amplitudes $x_i$ and $e_i$ can be reproduced without additional noise.
 
\medskip
 Next, we will discuss the impact of quantum noise on the performance of CIM. As indicated in Eqs.\\
  (\ref{eq:Wigner SDE 1})-(\ref{eq:error detection 2}), the relative magnitude of the quantum noise in the signal and error pulses is governed by the saturation parameter $g$. 
 When $g$ increases, the ratio between the normalized pulse amplitudes ($ x_i, e_i $) and normalized quantum noise amplitudes ($ gn_i, gm_i $) decreases. 
 Therefore, the CIM performance is expected to degrade as $g$ increases. However, as $g$ increases, the OPO threshold pump power decreases (see Figure C1 in \cite{Kako2020}), which suggests that the OPO energy cost to solution can be potentially reduced with increasing $g$.

\begin{figure}
	\centering
	\includegraphics[width=0.3\textwidth]{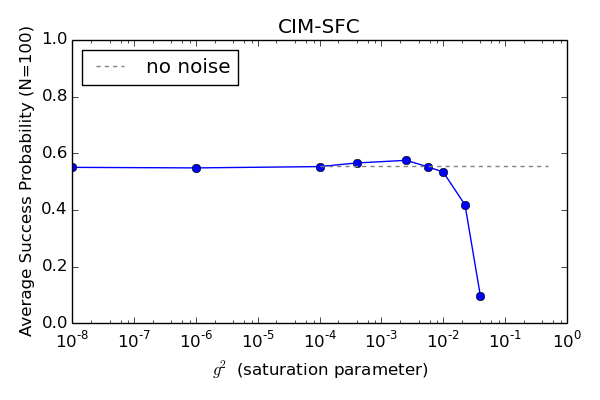} \includegraphics[width=0.3\textwidth]{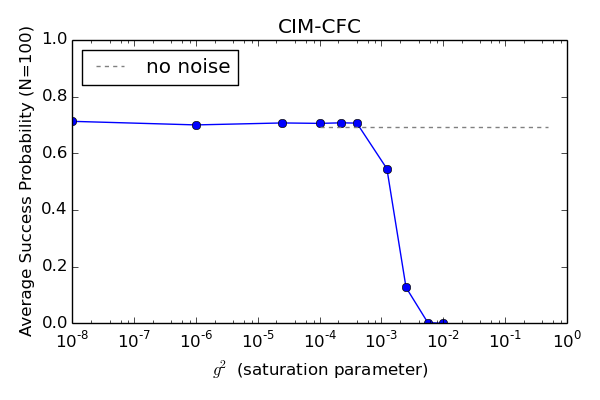}
	\caption{Success probability $ P_s $ vs. saturation parameter $ g^2 $ for CIM-SFC and CIM-CFC at 
	N=100. The success probability is averaged over 100 SK instances. 
	CIM-CAC is not shown; however, the result is nearly identical to that of CIM-CFC.}
	\label{fig:ps vs g2}
\end{figure}

\medskip
Figure \ref{fig:ps vs g2} shows the success probability  $ P_s $ for $ N=100 $ Ising problems (SK model) plotted against the saturation parameter $ g^2 $. The reflectivity of the extraction beam splitter $R_B$ is assumed to be $R_B = 0.1$. 
The success probability $ P_s $ is almost independent of the saturation parameter $ g^2 $ as long as $ g^2 \lesssim 10^{-4} $. However, when $ g^2 $ exceeds $ 10^{-3} $, the success probability drops rapidly owing to the decreased signal-to-quantum noise ratio, as mentioned above.

\medskip
Figure \ref{fig:ETS_noise_plot} shows the energy cost to solution for Ising problems (SK model) with $ N=100 $ and $ N=800 $, where we consider only the pump power to the main cavity PSA: $ E_{\small\textrm{main}} = 2 \hbar \omega \left( \textrm{MVM} \right) N \Delta t / g^{2}$, 
where MVM is the number of matrix-vector multiplication steps to solution and $ \Delta t $ is a round-trip time normalized by the signal lifetime ($\sim 0.1 $). 

\begin{figure}[!htb]
	\centering
	\includegraphics[width=0.3\textwidth]{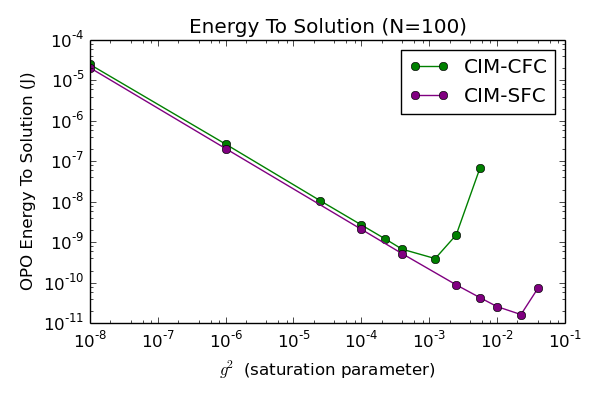} \includegraphics[width=0.3\textwidth]{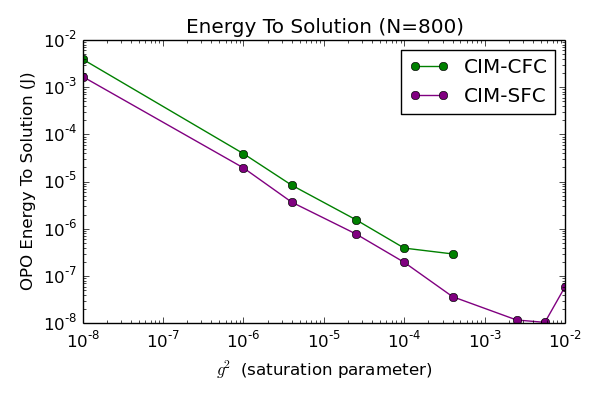}
	\caption{Energy cost to solution in joules of CIM-SFC and CIM-CFC considering only pump power to main cavity PSA. 
	The median ETS is plotted as a function of $g^2$ for N=100 and N=800 SK instances to show the optimal value of $g^2$ in each case.}
	\label{fig:ETS_noise_plot}
\end{figure}

\medskip
In Figure \ref{fig:ETS_noise_plot}, we can see that CIM-SFC is more robust to quantum noise compared to CIM-CFC, allowing us to potentially use a larger value of $g^2$. 
This is to be expected owing to the different roles payed by the error variable $e_i$ in each system. 
In CIM-CFC, the feedback signal is calculated as 

$$\tilde{z_i} = \tilde{e_i} \sum_j \xi  J_{ij} \tilde{x_j}(t)$$
which is the main cause of performance degradation when the quantum noise is increased. 
This is because, if the coherent excitation of $\tilde{e_i}$ is large, then small errors in $\sum_j   J_{ij}\xi \tilde{x_j}(t)$ 
will be amplified, and conversely, if the coherent excitation of $\sum_j \xi  J_{ij} \tilde{x_j}(t)$ is large, then small errors in $\tilde{e_i}$ will be amplified. 
There are no such beat noise components in CIM-SFC. Therefore, CIM-SFC is more robust to quantum noise. 
Moreover, the nonlinear function $\tanh(c \tilde{z_i})$ can help to suppress the quantum noise.

\medskip
Although they are not shown, the results for CIM-CAC are nearly identical to those for CIM-CFC.

\medskip
If we include the energy cost in the optical error correction circuit and pump pulse factory (as described in Figure \ref{fig:all-optical CIM}), the energy cost is increased by several orders of magnitude, as shown in Figure \ref{fig:Energy to solution}. 
Here, we assume that the pump pulse energy for a small signal amplification ($\sim$ 10 dB) in PSA$_1$, PSA$_2$, ... , PSA$_N$ and PSA$_e$ in the optical error correction circuit is 100 fJ/pulse, and that for a large signal amplification ($\sim$ 50 dB) in PSA$_0$ is 1 pJ/pulse.
These numbers correspond to the experimental values for a thin-film LiNO$_3$ ridge waveguide DOPO at a pump wavelength of 780 nm and a pump pulse duration of 100 fs\cite{TFLN}.
The pump energy consumed in the optical error correction circuit is estimated as $ E_{\small\textrm{correction}} = \left[ \left( N + 1 \right) \times 10^{-13} + 10^{-12} \right] N \left( \textrm{MVM} \right) (J) $.
The energy consumption in the pump pulse factory is attributed to three components: those of a 100-GHz soliton frequency comb generator, EOM modulators, and phase-sensitive amplifiers (Figure \ref{fig:all-optical CIM}(b)).
The 100-GHz soliton frequency comb generator requires an input power of $\sim$ 100 mW.{\scriptsize $^{\cite{Stern2018}}$}
The 100-GHz EOM modulators require an electrical input power of $\sim$ 400 mW each.{\scriptsize $^{\cite{Wang2018}}$}
The energy cost per pulse for PSA$_p$ is $\sim$ 1 pJ, while those for $N$ PSA$_s$ for the storage ring cavities are $\sim$ 100 fJ each. 
Note that $N$ EOM$_\textrm{s} $ (EOM$_1$, EOM$_2$, $\dots$ EOM$_N$) need to be operated only for one initial round-trip time, $10^{-11} N$ (sec).
The operational powers of the active devices in the 100-GHz CIM are summarized in Table \ref{tab:operational powers}.
The energy cost in the pump pulse factory is $ E_{\small\textrm{factory}} = \left[ 1.3 \times 10^{-11} N (\textrm{MVM}) + 4 \times10^{-12} N^{2} + \left( 10^{-12} + 10^{-13} N \right) (\textrm{MVM}) N \right] (J) $.
Table \ref{tab:energy costs} summarizes the energy costs in three parts of the CIM.

\begin{figure}[!htb]
	\centering
	\includegraphics[scale=0.4]{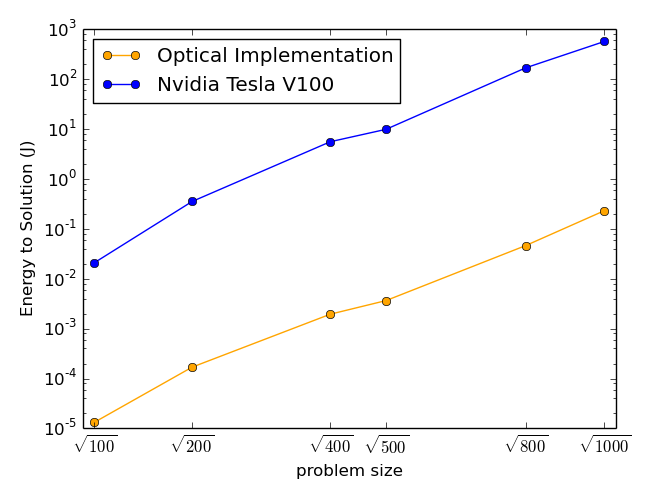}
	\caption{Estimated energy cost to solution of optical and GPU implementations of CIM-CAC vs. problem size $ \sqrt{N} $.
	The energy cost to solution for the optical CIM is based on the results presented in Table \ref{tab:energy costs}. 
	The energy cost of the GPU is based on the \~200W power consumption of the Nvidia Tesla V100 GPU used.}
	\label{fig:Energy to solution}
\end{figure}

\begin{table}
	\begin{center}
		\caption{Operational power of active photonic devices in 100 GHz CIM.}
		\label{tab:operational powers}
	\begin{tabular}{| l | r | c | }
	\hline 
	Devices & Power consumption & Reference\* \\ \hline
	Soliton frequency comb generator & 100 mW & \cite{Stern2018} \\
	Phase sensitive ampifier (PSA) 10 dB gain & 10 mW & \cite{TFLN} \\ 
	Phase sensitive ampifier (PSA) 50 dB gain & 100 mW & \cite{TFLN} \\ 
	EOM modulator & 400 mW & \cite{Wang2018} \\ \hline
	\end{tabular}
\end{center}
\end{table}

\begin{table}
	\begin{center}
		\caption{Energy cost to solution in three subsystems in CIM. MVM: matrix-vector multiplication steps to solution, $N$: problem size, one round trip time: $10^{-8}$ s, signal lifetime: $10^{-7}$ s.}
		\label{tab:energy costs}
		\begin{tabular}{| c || c | c | c |}
			\hline 
			Subsystem & Energy-to-solution\* \\ \hline
			Main cavity & $ 2.6 \times 10^{-20} (\textrm{MVM}) N / g^2 $  \\ \hline
			Optical error correction circuit & $ [ (N + 1) 10^{-13} + 10^{-12} ] (\textrm{MVM}) N \simeq 10^{-13} N^{2} (\textrm{MVM}) $ \\ \hline
			Pump pulse factory & $ [N \times 10^{-13} + 10^{-12} ] (\textrm{MVM}) N + 1.3 \times 10^{-11} N (\textrm{MVM}) + 4 \times 10^{-12} N^{2} \simeq 10^{-13} N^{2} (\textrm{MVM}) $ \\ \hline
		\end{tabular}
	\end{center}
\end{table}

\medskip
Figure \ref{fig:Energy to solution} shows the energy cost to solution if the CIM-CFC algorithm is implemented on GPU. The detailed description of this approach will be given in the next section. 
Even though the optical implementation of the error correction circuit and pump pulse factory as described in Figure \ref{fig:all-optical CIM} is technologically challenging, the energy cost can be decreased by several orders of magnitude compared to a modern GPU.

\section{CIM - Inspired Heuristic Algorithms}

\subsection{Scaling performance of CIM-CAC, CIM-CFC, and CIM-SFC}
To test whether the three classical nonlinear dynamics models given by Eqs. (\ref{eq:i-th spin 1})-(\ref{eq:third system 3}), are good Ising solvers, we can numerically integrate them on a digital platform. 
In this section, we will consider these CIM-inspired algorithms when numerically integrated using an Euler step. 
In addition, to ensure numerical stability, we constrain the range of some variables, the details of which are presented in Appendix A.

\medskip
The relevant performance metric is the time to solution or TTS (the number of integration time steps required to achieve a success rate of 99\%). 
In particular, we study how the median TTS scales as a function of the problem size for randomly generated SK spin glass instances (the couplings are chosen randomly between $+1$ and $-1$). 
The median TTS is computed on the basis of a set of 100 randomly generated instances per problem size, and 3200 trajectories are used per instance to evaluate the TTS.

\medskip
Figure \ref{fig:scaling_tts} shows the median TTSs of the three CIM-inspired algorithms (CIM-CAC, CIM-CFC, and CIM-SFC) are shown with respect to the problem size. The shaded regions represent 25th-75th percentiles. 
The linear behavior of the TTS with respect to $\sqrt{N}$ indicates that these algorithms have the same root exponential scaling of TTS that is also observed in physical CIMs with quantum noise from external reservoirs.{\scriptsize $^{\cite{Hamerly2019, Kako2020}}$} 
All three algorithms appear to have very similar scaling coefficients if the TTS is assumed to be of the form $TTS \approx A \cdot B^{\sqrt{n}}$. 
In addition to the similar scaling, all three algorithms show a similar spread (25th–75th percentile) in TTS, as indicated by the shaded region above. 
Although CIM-SFC may have a slightly larger spread in all cases, the spread does not appear to increase for larger problem sizes.

\begin{figure}[!htb]
	\centering
	\includegraphics[scale=0.4]{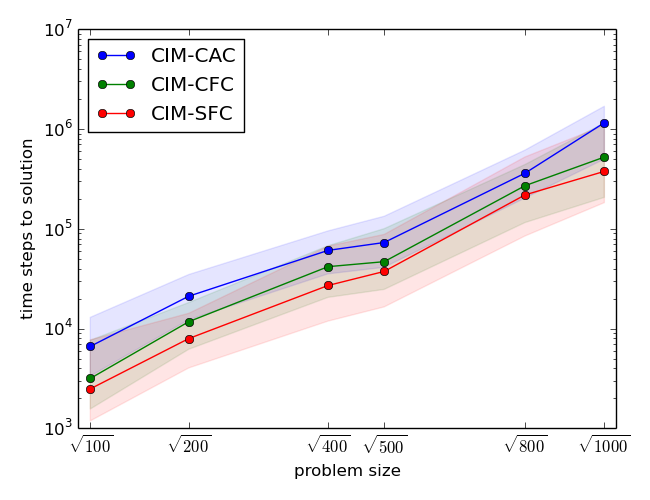} \includegraphics[scale=0.4]{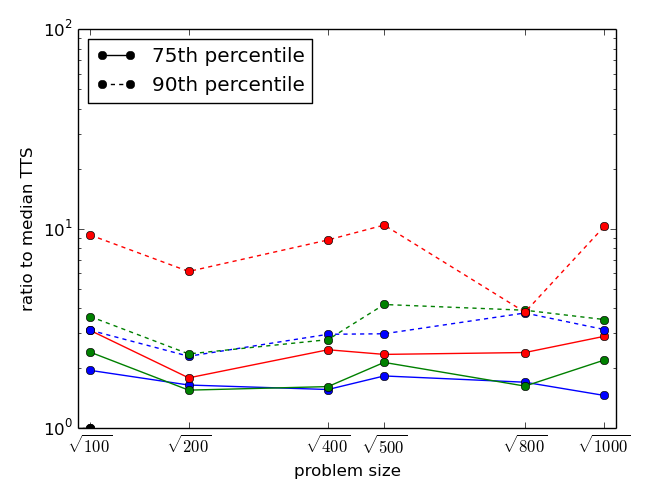}
	\caption{(left) TTSs of CIM-CAC, CIM-CFC and CIM-SFC vs. problem size $\sqrt{N}$. The shaded regions represent the 25th-75th percentile TTS. 
	(right) The 75th and 90th percentile TTS compared to the median TTS for all three systems as a function of the problem size.}
	\label{fig:scaling_tts}
\end{figure}

\subsection{Comparison with noisy mean field annealing (NMFA)} 
To show the importance of the auxiliary variable (error pulse) in CIM-SFC, we compared its performance to another CIM-inspired algorithm, namely noisy mean field annealing (NMFA).{\scriptsize $^{\cite{NMFA}}$} NMFA also applies a hyperbolic tangent function to the mutual coupling term. 
However, it does not have an auxiliary variable and relies on (artificial) quantum noise to escape from local minima. Figure \ref{fig:nmfa}, compares the scaling of NMFA to CIM-SFC with different values of the feedback parameter $k$. 
As $k$ controls the strength of the destabilization force caused by the auxiliary variable, we can measure the importance of the term $k(z_i - e_i)$ to the scaling behavior.
When $k = 0$, CIM-SFC is nearly identical to NMFA. The fact that CIM-SFC with $k=0$ shows slightly worse performance indicates that the noise included in NMFA likely has a small effect and may help destabilize the local minima (which can also be observed in Figure \ref{fig:ps vs g2}). 
The case $k=0.15$ is shown as an intermediate case, and $k=0.2$ is the (experimentally obtained) optimal value for $k$ in CIM-SFC. 

\medskip
As can be seen, the addition of the error correction feedback term $k(z_i -e_i)$ in Eq. (\ref{eq:third system 2}) is effective in improving both the scaling and the spread of TTS for the SK instances. 
This implies that the ``correlated artificial noise” provided by the auxiliary variable is more effective in finding better solutions than the “random quantum noise” from reservoirs.

\begin{figure}[!htb]
	\centering
	\includegraphics[scale=0.4]{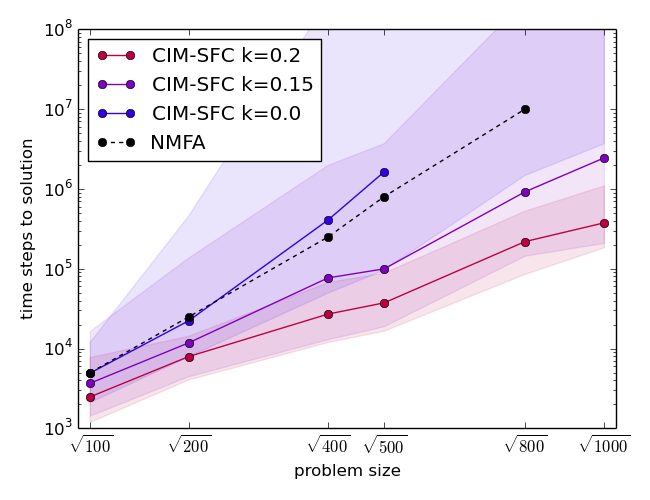} \includegraphics[scale=0.4]{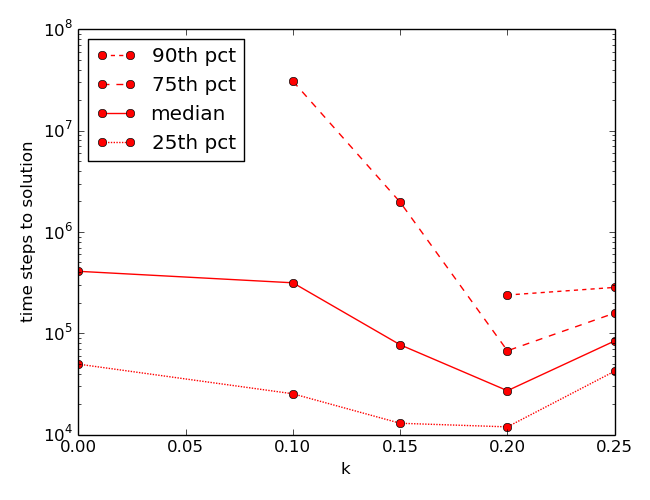} 
	\caption{(left) TTSs of CIM-SFC and NMFA vs. problem size $\sqrt{N}$. The shaded regions represent 25-75 percentile TTS. The results for NMFA are from \cite{Leleu2021}. 
	(right) TTS for N=400 SK instances for different values of $k$. The 75th and 90th percentiles are not shown for smaller values of $k$ because they were too large to be computed.}
	\label{fig:nmfa}
\end{figure}

\subsection{Comparison with discrete simulated bifurcation machine (dSBM)}
We compared the performance of the CIM-inspired algorithms with that of another heuristic Ising solver, namely the discrete simulated bifurcation machine (dSBM).{\scriptsize $^{\cite{Goto2019, Goto2021, Tatsumura2021}}$} 
Similar to CIM, dSBM also makes use of analog spins and continuous dynamics to solve combinatorial optimization problems.

\medskip
We are aware that the authors of \cite{Goto2021} seem to claim that dSBM is algorithmically superior to CIM-CAC by comparing the required number of MVMs to solution.
Although the authors of \cite{Goto2021} discussed the wall clock TTS of their implementations on many problem sets, when making the claim of algorithmic superiority, they only used the median TTS (in units of MVM) on SK instances for two problem sizes. 
In this section, we will provide a more detailed comparison of the three algorithms (CIM-CAC, CIM-CFC and CIM-SFC) with dSBM using MVM to solution (or equivalently, integration time steps to solution) as the performance metric. 
As mentioned before, this is a good comparison because all these algorithms will have MVM as the computational bottleneck when implemented on a digital platform. 
As discussed in Section 4, the computation of the Ising energy can be left until the end of the trajectory in most cases; thus, for this section, we will only consider the MVM involved in the computation of the mutual coupling term when calculating the MVM to solution. 

\medskip
The problem instance sets used in this section are: 
\begin{enumerate}
	\item A set of 100 randomly generated 800-spin SK instances (available upon request from the authors). This instance set contains fully connected instances with weights of $+1,-1$.
	\item The G-set instances that have been used as a benchmark for max-cut performance (available at https://web.stanford.edu/ yyye/yyye/Gset/). In this study, we consider 50 instances with a problem size of 800–2000. These instances have varying edge density and include weights of either $+1,0$ or $+1,0,-1$.
	\item Another set of 1000 randomly generated 800-spin and 1200-spin SK instance (available upon request from the authors) used to evaluate the worst-case performance.

\end{enumerate}

\medskip
To compare the performance on the 800-spin SK instances, the dSBM algorithm was also implemented on GPU. The parameters for dSBM were chosen on the basis of the parameters in \cite{Goto2021} (see Appendix D).

\begin{figure}[!htb]
	\centering
	\includegraphics[scale=0.4]{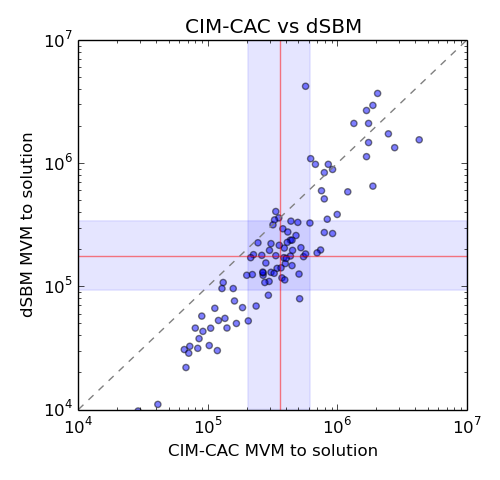}
	\includegraphics[scale=0.4]{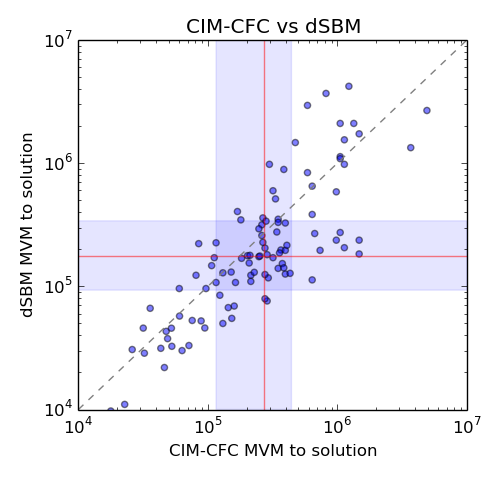}
	\includegraphics[scale=0.4]{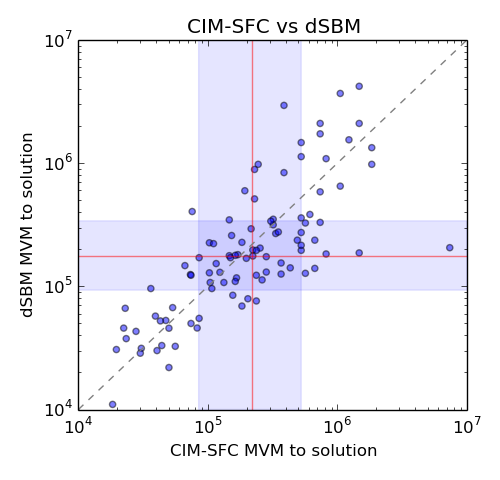}
	\caption{Required number of MVMs to solution for CIM-CAC, CIM-CFC, and CIM-SFC  vs. that for dSBM. The median TTS is indicated by the red lines and the 25th–75th percentiles are indicated by the shaded blue regions.}
	\label{fig:sk_tts}
\end{figure}

\medskip
In Figure \ref{fig:sk_tts}compares the performance of the three algorithms (CIM-CAC, CIM-CFC, and CIM-SFC) instance by instance with that of dSBM on the 800-spin SK instance set. The ground-state energies used
to evaluate the MVM to solution are the lowest energies found by the four algorithms.
As all four algorithms found the same lowest energies, it is highly likely that these are true ground-state energies. 
The parameters for all four systems can be found in Appendix A. 
As shown in Figure \ref{fig:sk_tts}, all four systems showed remarkably similar performance on the 800-spin instances when the parameters were optimized. 
It is important to note that with the parameters used in Figure \ref{fig:sk_tts}, CIM-SFC did not find the ground state in one instance. 
However, if different parameters are used, CIM-SFC will find the ground state for this particular instance as well. 
Thus, although CIM-SFC can achieve high performance, it is highly sensitive to parameter selection.

\medskip
The median TTS (in the units of MVM) of CIM-CFC, CIM-SFC and dSBM are nearly the same: around $2 \times 10^{5}$. Furthermore, the spread in TTS of these three algorithms is similar. 
Although CIM-CAC shows slightly worse median TTS (by less than a factor of two), it is worth noting that the instances in which CIM-CAC performs better than dSBM tend to be the harder instances. 
This indicates that among the four algorithms, CIM-CAC may show slightly better worst-case performance. 
We investigate this further later in this section. This pattern can also be seen on the G-set.

\medskip
Overall, all four algorithms show similar performance on the fully connected instance set, and it is impossible to determine which particular algorithm is the most effective one for this problem type. 
In addition to the similar median TTS and spread, there is a high level of correlation in the TTS among all four systems. 
This indicates either that instance difficulty is a universal property for all Ising heuristics or that there is something fundamentally common to the four algorithms. 
See Appendix E for a further discussion of the similarities and differences between these four systems.

\medskip
Although CIM-SFC shows good performance on fully connected problem instances, it struggles on many G-set instances. 
In Appendix D, we discuss a partial reason for this failure; however, the full reason is yet to be understood. 
In the future, we expect to modify CIM-SFC or find better parameters so that it can solve all problem types; however, for now, we will just consider CIM-SFC as a fully connected (or densely connected) Ising solver and compare only CIM-CAC, CIM-CFC, and dSBM on the G-set. 
For some results of CIM- SFC on the G-set, see Appendix D.

\begin{figure}[!htb]
	\centering
	\includegraphics[scale=0.3]{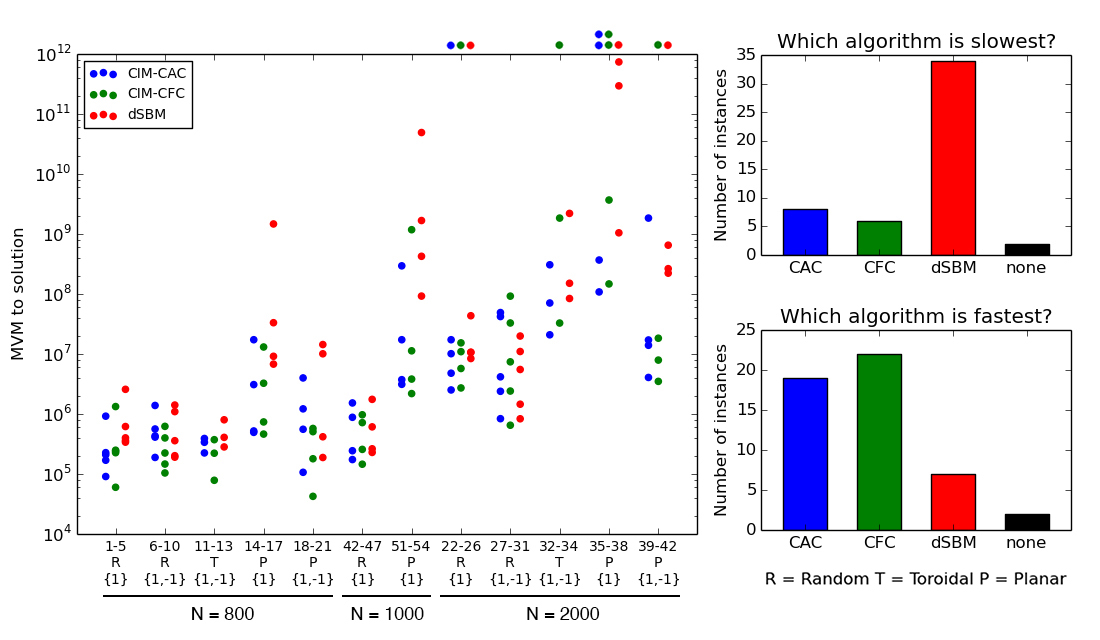}

	\caption{(left) TTS on G-set graphs for CIM-CAC, CIM-CFC
		and dSBM. The best known cut values are found in \cite{Goto2021}.
		The TTS for dSBM is from \cite{Goto2021}. In this plot, the instances are separated into groups depending on the graph type and size. 
		The dots above the plot indicate that the best known cut value was not found.
		(right) Histograms showing which algorithm realizes the slowest TTS and which algorithm has the fastest TTS. 
		The column labeled ``none" indicates the two instances in which none of the three algorithms found the best known cut value.
		The parameters are chosen and optimized separately for each instance type. An instance-by-instance comparison and the parameters used are presented in Appendix B.
	}
	\label{fig:gset_tts}
\end{figure}

\medskip
All three algorithms (CIM-CAC, CIM-CFC, and dSBM) show fairly good performance on the G-set; however, we argue that CIM-CAC is the most consistently effective algorithm. 
CIM-CAC and dSBM were able to find the best known cut values in 47 out of 50 instances, while CIM-CFC found the best known cut value in 45 out of 50 instances. 
It is worth noting that the simulation time used to calculate the TTS for dSBM {\scriptsize $^{\cite{Goto2021}}$} was much longer than that used in this study. Given the same simulation time, dSBM would most likely have solved only 45 out of 50 instances. 
As shown in Figure \ref{fig:gset_tts}, CIM-CAC and CIM-CFC are faster (in units of MVM) than dSBM in most instances.
More importantly, among the instances in which dSBM is faster, there are no cases where dSBM is significantly faster than CIM-CAC, other than G37, in which CIM-CAC did not find the best known cut value. Meanwhile, we found that CIM-CAC was more than an order of magnitude faster than dSBM in 13 out 50 instances. Therefore, we believe that CIM-CAC is a more reliable algorithm when considering many problem types.

\medskip
The difference between CIM-CAC and CIM-CFC is subtle. 
This is to be expected, as the dynamics of the two systems are very similar. 
Although the performance of the two algorithms for G-set is nearly identical in most cases, for some of the harder instances, there are some cases in which CIM-CFC cannot find the best known cut value or CIM-CFC has a significantly longer TTS. 
This indicates that CIM- CAC is fundamentally a more promising algorithm or that precise parameter selection for CIM-CFC is required.

\begin{figure}[!htb]
	\centering
	\includegraphics[scale=0.4]{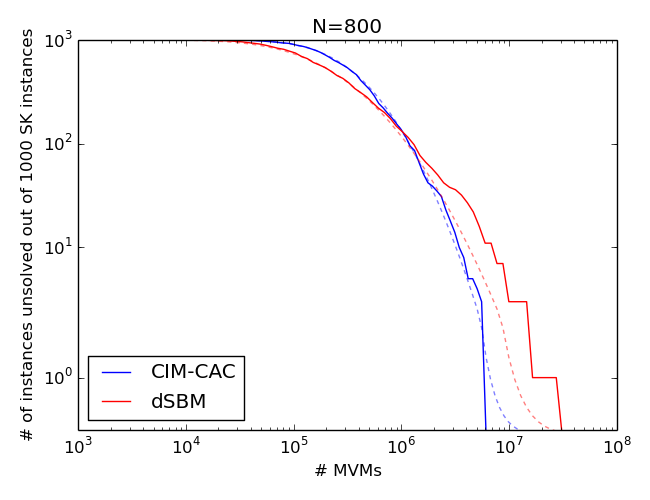}
	\includegraphics[scale=0.4]{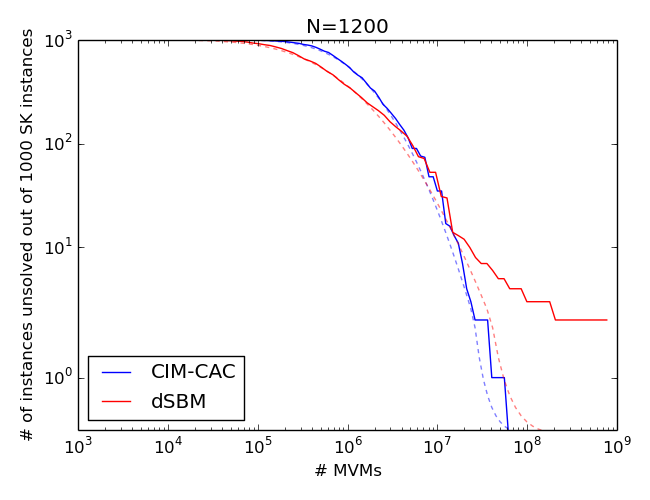}
	\caption{Number of instances that remain unsolved (success probability under 99\%) after a certain number of MVMs (time steps) for CIM-CAC and dSBM. The dotted lines represent the assumption of a log-normal distribution for the TTS on randomly generated SK instances.
	The instance sets used are 1000 randomly generated SK instances (different from the 100 instances used in Figure \ref{fig:sk_tts}) of problem sizes N=800 (top) and N=1200 (bottom). The parameters for both systems can be found in Appendix A.
	}
	\label{fig:CAC-SBM comparison}
\end{figure}

\medskip
As noted in Figure \ref{fig:sk_tts}, the worst case performance of CIM-CAC may be slightly better than that of dSBM. To evaluate this further we created new sets of 1000 800-spin and 1200-spin SK instances. 
Figure \ref{fig:CAC-SBM comparison} shows the number of instances solved as a function of the number of MVMs required to achieve a success probability of 99\%. As can be seen in both cases, dSBM can solve the easier instances with fewer MVMs; however, for the hardest instances, CIM-CAC is faster.
his can be understood by observing the intersection point of the two curves in Figure \ref{fig:CAC-SBM comparison}.

\medskip
In nearly all cases, the best Ising energy found was the same for both the solvers when a similar number of MVMs were used (see Appendix A for the parameters). 
However, for two instances in the N=1200 set, the Ising energy found by CAC was not found by dSBM. This remained true even when 50,000 dSBM trajectories were used for these instances.

\medskip
Our results suggest that dSBM may struggle considerably for some harder SK instances. 
However, we acknowledge that this could be a result of sub-optimal parameter selection for dSBM. The parameters used (see appendix A) were optimized manually to achieve a good median TTS;
however, they may not be the best parameters if one wants to solve the hardest instances. 
By contrast, for CIM-CAC, the optimal parameters for the median TTS appear to also perform well on the hardest instances.

\medskip
To ensure that an Ising solver can find the true ground state of a given problem, the worst-case performance is very important. For this purpose, we believe that CIM-CAC is likely the more fundamentally superior algorithm, at least in the case of randomly generated SK instances. For the other CIM modifications, this is likely not true. In particular, for CIM-SFC, the worst-case performance is significantly worse than that of dSBM and CIM-CAC (as shown in Figure \ref{fig:scaling_tts}). 
In the future, it would be interesting to investigate the cause for this phenomenon and also to examine the spread of TTS for different problem types.

\section{Conclusion}
The new coherent Ising machines presented in this paper (CIM-CFC and CIM-SFC) have considerable potential as both digital heuristic algorithms and optically implemented physical devices. 
Rapid advances in the thin-film LiNbO$_3$ platform{\scriptsize $^{\cite{TFLN,Stern2018,Wang2018}}$} as a photonic integrated circuit technology might enable the proposed optical CIM to surpass existing digital algorithms on a CMOS platform in terms of both speed and energy consumption.

\medskip
The proposed CIM-inspired algorithms were shown to be fast and accurate Ising solvers even when implemented on an existing digital platform. 
In particular, we showed that their performance is very similar to that of other existing analog-system-based algorithms such as dSBM.
This again brings up the question raised in \cite{Leleu2019} as to whether the simulation of analog spins on a digital computer can outperform a purely discrete heuristic algorithm. 
Finally, whether chaotic dynamics are necessary for a deterministic dynamical system to be a good Ising solver is left as an open issue for future research.{\scriptsize $^{\cite{Leleu2019,Leleu2021,Goto2021}}$}

\section*{Appendix A: Optimization of simulation parameters}
Here we summarize the simulation parameters used in our numerical experiments. 
The parameters are optimized empirically and thus do not necessarily reflect the true optimum values.
\subsection*{Parameters used in Figure \ref{fig:trajectories}}
\begin{center}
\begin{tabular}{ c| c }
& CIM-SFC (upper left panel) \\ \hline
 N step & 500 \\ 
 \hline
 $\Delta T$ & 0.4 \\  \hline
$p$ & -1.0  \\ \hline
$c$ & 1.0  \\ \hline
$\beta$ & 0.3  \\ \hline
$k$ & 0.2 

\end{tabular}
\quad
\begin{tabular}{ c| c }
& CIM-CFC (upper right panel) \\ \hline
 N step & 1000 \\ 
 \hline
 $\Delta T$ & 0.4 \\  \hline
$p$ & -1.0 \\ \hline
$\alpha$ & 1.0 \\ \hline
$\beta$ & 0.2 \\ \hline

\end{tabular}
\begin{tabular}{ c| c }
& CIM-SFC (lower left panel) \\ \hline
 N step & 500 \\ 
 \hline
 $\Delta T$ & 0.4 \\  \hline
$p$ & -1.0 $\rightarrow$ 1.0 \\ \hline
$c$ & 1.0 $\rightarrow$ 3.0 \\ \hline
$\beta$ & 0.3 $\rightarrow$ 0.1 \\ \hline
$k$ & 0.2 

\end{tabular}
\quad
\begin{tabular}{ c| c }
& CIM-CFC (lower right panel) \\ \hline
 N step & 1000 \\ 
 \hline
 $\Delta T$ & 0.4 \\  \hline
$T_r$ & 900 \\ \hline
$T_p$ & 100 \\ \hline
$p$ & -1.0 $\rightarrow$ 1.0 \\ \hline
$\alpha$ & 1.0 \\ \hline
$\beta$ & 0.2 \\ \hline

\end{tabular}

\end{center}

\medskip
\subsection*{Parameters used in Figure \ref{fig:scaling_tts}, \ref{fig:nmfa}, and \ref{fig:sk_tts}}

\subsubsection*{CIM-CAC}
In our simulation, the $x_i$ variables are restricted to the range $[ -\frac{3}{2}\sqrt{\alpha} , \frac{3}{2}\sqrt{\alpha}]$ at each time step. 
The parameters $p$ and $\alpha$ are modulated linearly from their starting to ending values during the $T_r$ time steps and are kept at the final value for an additional $T_p$ time steps. 
The initial value $x_i$ is set to a random value chosen from a zero-mean Gaussian distribution with a standard deviation of $10^{-4}$ and $e_i = 1$. 
Furthermore, 3200 trajectories are computed per instance to evaluate TTS. The actual parameters used for simulation are listed below:

\begin{center}
	\begin{tabular}{ c| c }
		N step & 3200 \\ 
		\hline
		$\Delta T$ & 0.125 \\  \hline
		$T_r$ & 2880 \\ \hline
		$T_p$ & 320 \\ \hline
		$p$ & -1.0 $\rightarrow$ 1.0 \\ \hline
		$\alpha$ & 1.0 $\rightarrow$ 2.5 \\ \hline
		$\beta$ & 0.8 \\ \hline	
	\end{tabular}
\end{center}

\medskip

\subsubsection*{CIM-CFC}
In our simulation the $x_i$ variables are restricted to the range $[ -1.5, 1.5]$ and $e_i$ is restricted to the range $[0.01, \infty]$. 
The parameter $p$ is modulated linearly from its starting to ending values during the first $T_r$ time steps and kept at the final value for an additional $T_p$ time steps. 
The initial value $x_i$ is set to a random value chosen from a zero-mean Gaussian distribution with a standard deviation of $0.1$ and $e_i = 1$. 
Furthermore, 3200 trajectories are computed per instance to evaluate TTS. The actual parameters used for simulation are listed below:

\begin{center}
	\begin{tabular}{ c| c }
		N step & 1000 \\ 
		\hline
		$\Delta T$ & 0.4 \\  \hline
		$T_r$ & 900 \\ \hline
		$T_p$ & 100 \\ \hline
		$p$ & -1.0 $\rightarrow$ 1.0 \\ \hline
		$\alpha$ & 1.0 \\ \hline
		$\beta$ & 0.2 \\ \hline
		
	\end{tabular}
\end{center}

\subsubsection*{CIM-SFC}
Restriction of $x_i$ and $e_i$ variables is not needed as this system is more numerically stable. The parameters $p$, $c$ and $\beta$ are modulated linearly from their starting to ending values during simulation. The initial value $x_i$ is set to a random value chosen from a zero-mean Gaussian distribution with standard deviation of $0.1$ and $e_i = 0$. 
3200 trajectories are computed per instance to evaluate TTS. Actual parameters used for simulation are listed below:

\begin{center}
	\begin{tabular}{ c| c }
		N step & 500 \\ 
		\hline
		$\Delta T$ & 0.4 \\  \hline
		$p$ & -1.0 $\rightarrow$ 1.0 \\ \hline
		$c$ & 1.0 $\rightarrow$ 3.0 \\ \hline
		$\beta$ & 0.3 $\rightarrow$ 0.1 \\ \hline
		$k$ & 0.2 
		
	\end{tabular}
\end{center}

\medskip
In addition to the above-mentioned parameters, it is important for the normalizing factor $\xi$ for the mutual coupling term to be chosen as \cite{Kako2020},
\begin{eqnarray}
	\xi = \sqrt{\frac{2N}{\sum J_{ij}^2}}. \nonumber
\end{eqnarray}

\medskip
This choice is crucial for the successful performance of CIM-SFC but not for CIM-CAC and CIM-CFC.

\medskip
Moreover, it is important to note that we used the same number of time steps for all the problem sizes
in Figures \ref{fig:scaling_tts} and \ref{fig:nmfa}.
It is likely that the optimal number of time steps is smaller for smaller problem sizes, thus the scaling of TTS when the number of time steps is optimized separately for each problem size might 
be slightly worse than the reported scaling. 
However, we do not believe that this difference would be very significant. For the scaling of TTS for CIM-CAC when different parameters are chosen, see Appendix C.

\medskip
\subsubsection*{dSBM}
For Figure \ref{fig:sk_tts}, dSBM is implemented as described in \cite{Goto2021}. The parameters used are

\begin{center}
	\begin{tabular}{ c| c }
		N step & 2000 \\ 
		\hline
		$\Delta T$ & 1.25 \\  \hline
		$c$ & 0.5 \\
	\end{tabular}
\end{center}

\subsection*{Parameters used in Figure \ref{fig:CAC-SBM comparison}}
The parameters for N=800 are the same as those for Figure \ref{fig:sk_tts}. The parameters for N=1200 are listed below. 
The number of trajectories used for N=1200 was 3200 for most instances; however, to accurately evaluate the success probability, 10000-50000 trajectories were computed for the 10 hardest instances for both algorithms. 
Moreover, owing to the hardness in the case of N=1200, we are not very certain that the true ground state was found.

\subsubsection*{CIM-CAC}
\begin{center}
	\begin{tabular}{ c| c }
		N step & 8000 \\ 
		\hline
		$\Delta T$ & 0.125 \\  \hline
		$T_r$ & 7200 \\ \hline
		$T_p$ & 800 \\ \hline
		$p$ & -1.0 $\rightarrow$ 1.0 \\ \hline
		$\alpha$ & 1.0 $\rightarrow$ 2.5 \\ \hline
		$\beta$ & 0.8 \\ \hline	
	\end{tabular}
\end{center}

\subsubsection*{dSBM}
\begin{center}
	\begin{tabular}{ c| c }
		N step & 4000 \\ 
		\hline
		$\Delta T$ & 1.25 \\  \hline
		$c$ & 0.5 \\
	\end{tabular}
\end{center}

\subsection*{Numerical Integration}
An Euler step is used for integration in all the cases (except for dSBM). As described above we constrain the range of $x_i$ variables to ensure numerical stability.
This is not necessary for performance but allows us to increase the integration time step by a factor of 2 or 3 without compromising the success probability. 
In Figure \ref{fig:time_step_comparison} we show the success probability of CIM-CAC with respect to the time step for both constrained and unconstrained systems.

\medskip
The results in Section 5 for CIM-CFC do not use this numerical constraint the CIM in Section 5 is meant to be a physical machine, and a time step of 0.2 is used.
\section*{Appendix B: Simulation Results for G-set}

The results in Figure \ref{fig:gset_tts} for dSBM are taken directly from the GPU implementation of dSBM in \cite{Goto2021}. The unit for TTS in Table \ref{tab:TTS} is time steps to solution, or equivalently, MVM to solution. 
In our simulation, 3200, 10000, or 32000 trajectories were generated to evaluate the TTS depending on the instance difficulty. 
The numbers in bold denote the best TTS among the three algorithms.

\begin{figure}[!htb]
	\centering
	\includegraphics[scale=0.5]{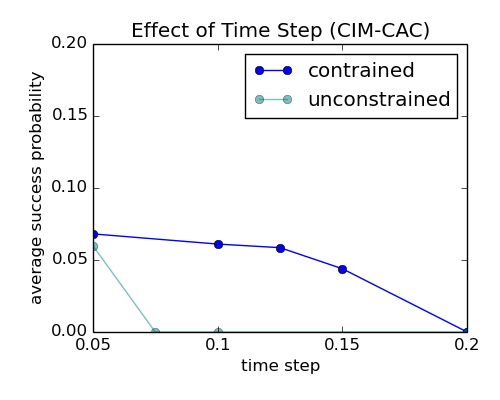}
	\caption{Success probability of CIM-CAC with respect to time step for both constrained and unconstrained systems. For the blue curve, the $x_i$ amplitudes are restricted to the range $[-1.5\sqrt{\alpha}, 1.5\sqrt{\alpha}]$}
	\label{fig:time_step_comparison}
\end{figure}

\medskip
\subsection*{CIM-CAC parameters for G-set}
The variables are restricted as described in Appendix A, and the initial conditions are set in the same way. The following parameters are the same for all the G-set instances.

\medskip
\begin{center}
	\begin{tabular}{ c| c }
		$\alpha$ & 1.0 $\rightarrow$ 3.0 \\ \hline
		$\beta$ & 0.3 \\
	\end{tabular}
\end{center}
The parameters $p$, $\Delta T$, and the number of time steps used in each phase are chosen by instance type as follows:
\begin{center}
	\begin{tabular}{| c | c | c | c | c | c | c | c | c |}
		\hline 
		Graph Type & Edge Weight & N & Instance \# & $p$ & N step & $\Delta T$ & $T_r$ & $T_p$ \\ \hline
		Random & \{+1\} & 800 &  1-5 & -0.5 $\rightarrow$ 1.0 & 6666 & 0.075 & 6000 & 666 \\  \hline
		Random & \{+1, -1\} & 800 & 6-10 & -0.5 $\rightarrow$ 1.0 & 6666 & 0.075 & 6000 & 666 \\  \hline
		Toroidal & \{+1, -1\} & 800 &  11-13 & -4.0 & 5000 & 0.1 & 4500 & 500 \\  \hline
		Planar & \{+1\} & 800 &  14-17 & -1.0 & 20000 & 0.05 & 18000 & 2000 \\  \hline
		Planar & \{+1, -1\} & 800 &  18-21 & -1.0 & 20000 & 0.05 & 18000 & 2000 \\  \hline
		Random & \{+1\} & 1000 &  43-46 & -0.5 $\rightarrow$ 1.0 & 10000 & 0.1 & 9000 & 1000 \\  \hline
		Planar & \{+1\} & 1000 &  51-54 & -1.0  & 20000 & 0.05 & 18000 & 2000 \\  \hline
		Random & \{+1\} & 2000 &  22-26 & -0.5 $\rightarrow$ 1.0 & 20000 & 0.1 & 19000 & 1000 \\  \hline
		Random & \{+1, -1\} & 2000 &  27-31 & -0.5 $\rightarrow$ 1.0 & 20000 & 0.1 & 19000 & 1000 \\  \hline
		Toroidal & \{+1, -1\} & 2000 &  32-34 & -4.0 $\rightarrow$ -3.0 & 20000 & 0.1 & 19000 & 1000 \\  \hline
		Planar & \{+1\} & 2000 &  35-38 & -1.0 $\rightarrow$ -0.5 & 80000 & 0.05 & 78000 & 2000 \\  \hline
		Planar & \{+1\} & 2000 &  39-42 & -1.0 $\rightarrow$ -0.5 & 80000 & 0.05 & 78000 & 2000 \\  \hline
	\end{tabular}
\end{center}

\subsection*{CIM-CFC parameters for G-set}
\medskip
The variables are restricted as described in Appendix A, and the initial conditions are set in the same way. The following parameters are the same for all the G-set instances.

\begin{center}
	\begin{tabular}{ c| c }
		$\alpha$ & 1.0 \\ \hline
		$\beta$ & 0.15 \\
	\end{tabular}
\end{center}
The parameters $p$, $\Delta T$, and the number of time steps used in each phase are chosen by instance type as follows:
\begin{center}
	\begin{tabular}{| c | c | c | c | c | c | c | c | c |}
		\hline 
		Graph Type & Edge Weight & N & Instance \# & $p$ & N step & $\Delta T$ & $T_r$ & $T_p$ \\ \hline
		Random & \{+1\} & 800 &  1-5 & -1.0 $\rightarrow$ 1.0 & 4000 & 0.125 & 3600 & 400 \\  \hline
		Random & \{+1, -1\} & 800 & 6-10 & -1.0 $\rightarrow$ 1.0 & 2000 & 0.25 & 1800 & 200 \\  \hline
		Toroidal & \{+1, -1\} & 800 &  11-13 & -3.0 $\rightarrow$ -1.0 & 2000 & 0.25 & 1800 & 200 \\  \hline
		Planar & \{+1\} & 800 &  14-17 & -2.0  $\rightarrow$ 0.0 & 8000 & 0.125 & 7200 & 800 \\  \hline
		Planar & \{+1, -1\} & 800 &  18-21 & -2.0  $\rightarrow$ 0.0 & 4000 & 0.25 & 3600 & 400 \\  \hline
		Random & \{+1\} & 1000 &  43-46 & -1.0 $\rightarrow$ 1.0 & 5000 & 0.2 & 4500 & 500 \\  \hline
		Planar & \{+1\} & 1000 &  51-54 & -2.0 $\rightarrow$ 0.0 & 16000 & 0.125 & 15200 & 800 \\  \hline
		Random & \{+1\} & 2000 &  22-26 & -1.0 $\rightarrow$ 1.0 & 10000 & 0.2 & 9500 & 500 \\  \hline
		Random & \{+1, -1\} & 2000 &  27-31 & -1.0 $\rightarrow$ 1.0 & 10000 & 0.2 & 9500 & 500 \\  \hline
		Toroidal & \{+1, -1\} & 2000 &  32-34 & -3.0 $\rightarrow$ -1.0 & 40000 & 0.1 & 39000 & 1000 \\  \hline
		Planar & \{+1\} & 2000 &  35-38 & -2.0 $\rightarrow$ 0.0 & 80000 & 0.05 & 78000 & 2000 \\  \hline
		Planar & \{+1\} & 2000 &  39-42 & -2.0 $\rightarrow$ 0.0 & 40000 & 0.1 & 39000 & 1000 \\  \hline
	\end{tabular}
\end{center}
\subsection*{Results on G-set}
Success probability and TTS of CIM-CAC, CIM-CFC and dSBM on G-set graphs. The success probability for CIM-CAC and CIM-CFC is for a single trajectory.
The results for dSBM are from the GPU implementation in \cite{Goto2021}.
\begin{center}
\begin{tabular}{| c | c | c | c | c | c | c |}
	\hline 
	Instance & CIM-CAC TTS & CIM-CAC $P_s$ &  CIM-CFC TTS & CIM-CFC $P_s$ &  dSBM TTS & dSBM $P_s$* \\ \hline
	G1 & 90805 & 0.286875 & \textbf{60078} & 0.264062 & 339332 & 0.987\\ \hline 
	G2 & \textbf{920217} & 0.0328125 & 1330454 & 0.01375 & 2578124 & 0.82\\ \hline 
	G3 & \textbf{169551} & 0.165625 & 249212 & 0.07125 & 400343 & 0.996\\ \hline 
	G4 & \textbf{209086} & 0.136563 & 239389 & 0.0740625 & 361673 & 0.983\\ \hline 
	G5 & 226881 & 0.126562 & \textbf{226449} & 0.078125 & 618221 & 0.972\\ \hline 
	G6 & 188908 & 0.15 & \textbf{104083} & 0.0846875 & 190728 & 0.979\\ \hline 
	G7 & 562413 & 0.053125 & \textbf{146490} & 0.0609375 & 201889 & 0.974\\ \hline 
	G8 & 431029 & 0.06875 & 399118 & 0.0228125 & \textbf{358947} & 0.954\\ \hline 
	G9 & 411604 & 0.071875 & \textbf{223836} & 0.0403125 & 1095704 & 0.867\\ \hline 
	G10 & 1388073 & 0.021875 & \textbf{622470} & 0.0146875 & 1410031 & 0.407\\ \hline 
	G11 & 337563 & 0.0659375 & \textbf{222079} & 0.040625 & 282524 & 0.98\\ \hline 
	G12 & 224452 & 0.0975 & \textbf{78562} & 0.110625 & 407997 & 0.973\\ \hline 
	G13 & 391011 & 0.0571875 & \textbf{373236} & 0.024375 & 800687 & 0.996\\ \hline 
	G14 & 17291018 & 0.0053125 & \textbf{13080721} & 0.0028125 & 1469967245 & 0.005\\ \hline 
	G15 & 521572 & 0.161875 & \textbf{462528} & 0.0765625 & 6782113 & 0.804\\ \hline 
	G16 & \textbf{494303} & 0.17 & 737146 & 0.04875 & 9156329 & 0.992\\ \hline 
	G17 & \textbf{3089154} & 0.029375 & 3256332 & 0.01125 & 33222397 & 0.283\\ \hline 
	G18 & 556635 & 0.1525 & \textbf{507805} & 0.035625 & 14375986 & 0.074\\ \hline 
	G19 & 1218307 & 0.0728125 & \textbf{179562} & 0.0975 & 417204 & 0.995\\ \hline 
	G20 & 106718 & 0.578125 & \textbf{42428} & 0.352187 & 188349 & 0.98\\ \hline 
	G21 & 3991180 & 0.0228125 & \textbf{574365} & 0.0315625 & 10080921 & 0.136\\ \hline 
	G43 & 174031 & 0.2325 & \textbf{145625} & 0.14625 & 228908 & 0.992\\ \hline 
	G44 & \textbf{244188} & 0.171875 & 257230 & 0.085625 & 263171 & 0.985\\ \hline 
	G45 & \textbf{880856} & 0.0509375 & 970877 & 0.0234375 & 1754473 & 0.985\\ \hline 
	G46 & 1528073 & 0.0296875 & 717957 & 0.0315625 & \textbf{610421} & 0.992\\ \hline 
	G51 & 3732360 & 0.024375 & \textbf{2189016} & 0.0331 & 424989992 & 0.067\\ \hline 
	G52 & \textbf{3122871} & 0.0290625 & 3820774 & 0.0191 & 92285270 & 0.213\\ \hline 
	G53 & 17291018 & 0.0053125 & \textbf{11298922} & 0.0065 & 1676440469 & 0.043\\ \hline 
	G54 & \textbf{294684837} & 0.0003125 & 1178886725 & 6.25e-05 & 49107077298 & 0.0006\\ \hline 
	G22 & \textbf{2516544} & 0.0359375 & 2718081 & 0.0168 & 8401394 & 0.928\\ \hline 
	G23 & N/A & 0.0 & N/A & 0.0 & N/A & 0.0\\ \hline 
	G24 & \textbf{4785454} & 0.0190625 & 5733406 & 0.008 & 10585339 & 0.648\\ \hline 
	G25 & 17291018 & 0.0053125 & \textbf{15327529} & 0.003 & 43414254 & 0.399\\ \hline 
	G26 & \textbf{10117012} & 0.0090625 & 10941648 & 0.0042 & 10730290 & 0.643\\ \hline 
	G27 & 835530 & 0.104375 & \textbf{649972} & 0.0684 & 832465 & 0.971\\ \hline 
	G28 & 2389444 & 0.0378125 & 2413498 & 0.0189 & \textbf{1455914} & 0.952\\ \hline 
	G29 & \textbf{4164219} & 0.021875 & 7404644 & 0.0062 & 5516820 & 0.737\\ \hline 
	G30 & 42058344 & 0.0021875 & 32871041 & 0.0014 & \textbf{11002259} & 0.738\\ \hline 
	G31 & 49075749 & 0.001875 & 92080375 & 0.0005 & \textbf{19923732} & 0.199\\ \hline 
	G32 & \textbf{70802710} & 0.0013 & 1841975969 & 0.0001 & 150969342 & 0.093\\ \hline 
	G33 & \textbf{306965291} & 0.0003 & N/A & 0.0 & 2204950868 & 0.005\\ \hline 
	G34 & \textbf{20886506} & 0.0044 & 32801883 & 0.0056 & 84156149 & 0.231\\ \hline 
	G35 & N/A & 0.0 & N/A & 0.0 & N/A & 0.0\\ \hline 
	G36 & \textbf{368321503} & 0.0005 & 3683951938 & 0.0001 & 736790387782 & 0.0001\\ \hline 
	G37 & N/A & 0.0 & N/A & 0.0 & \textbf{294701417831} & 0.0002\\ \hline 
	G38 & \textbf{108264816} & 0.0017 & 147181162 & 0.0025 & 1046295719 & 0.068\\ \hline 
	G39 & 4065368 & 0.0443 & \textbf{3497864} & 0.0513 & 651087484 & 0.107\\ \hline 
	G40 & 1841975969 & 0.0001 & N/A & 0.0 & \textbf{264354894} & 0.154\\ \hline 
	G41 & 17123320 & 0.0107 & \textbf{7916531} & 0.023 & 222414431 & 0.282\\ \hline 
	G42 & \textbf{13969282} & 0.0131 & 18328423 & 0.01 & N/A & 0.0\\ \hline 
\end{tabular}
\end{center}

*Note that $P_s$ for dSBM taken from \cite{Goto2021} is the success probability for a batch of 160 trajectories. To calculate the $P_s$ for a single dSBM trajectory use $1 - (1-P_s)^{\frac{1}{160}}$.

\section*{Appendix C: Reasoning for parameter selection}
The parameters are selected numerically for the most part; however, the choice of $p$, $\alpha$, and $\beta$ can be understood as follows. It is observed that the average residual energy visited by CIM-CAC during the search process can be roughly estimated by the formula.{\scriptsize $^{\cite{Leleu2019}}$}

\begin{eqnarray}
	\Delta E_{\textrm{avg}} \approx K \frac{1 - p}{\alpha \beta}  \nonumber
\end{eqnarray}

where K is a constant depending only on the problem type and size. This formula essentially predicts the effective sampling temperature of the system (although the distribution may not be an exact Boltzmann distribution). Based on this philosophy, we gradually reduce the ``system temperature” to produce an annealing effect.
This is the motivation for increasing $p$ and $\alpha$. The different choices for the range of $p$ on different G-set instances reflects the vastly different values for the constant $K$ depending on the structure of the max-cut problem. 

In a more general setting, the value of $K$ can be predicted on the basis of the problem type; thus, the range for $p$ and $\alpha$ can be chosen accordingly. 

\medskip
Although it has not been
verified, a similar formula most likely holds for CIM-CFC; thus, the parameters for CIM-CFC are chosen in the same way.

\subsection*{Optimal parameters with respect to problem size (CIM-CAC)}
\medskip

\begin{figure}[!htb]
	\centering
	\includegraphics[scale=0.4]{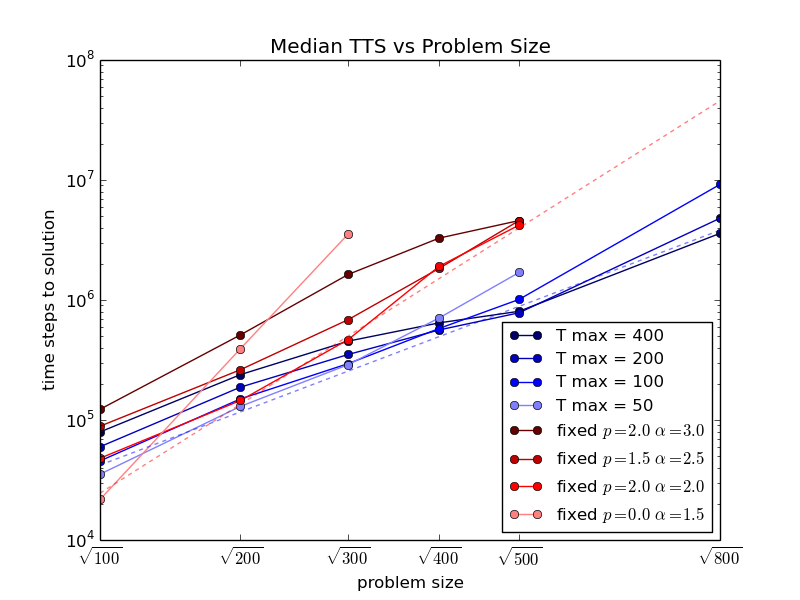}
	\caption{Performance of CIM-CAC with respect to problem size for different parameters. The fixed parameters are indicated in red while the parameter modulation is indicated in blue. The red and blue dotted lines are fits for the lower envelopes of the red and blue curves, respectively}
	\label{fig:scaling_params}
\end{figure}

Figure \ref{fig:scaling_params} shows the difference in scaling when the parameters are fixed (red shades) compared to when the parameters are modulated linearly (blue shades). 
In addition, the optimal annealing time (in other words, the optimal speed of modulation) will change with respect to the problem size, as we need long annealing times to get good results for large problem sizes. 
This pattern was also used when choosing parameters on the G-set.

\medskip
Although Figure \ref{fig:scaling_params} is based on the Gaussian quantum model for CIM-CAC in MFB-CIM (see \cite{Kako2020}), the difference in performance between this model and the noiseless model discussed in this paper is insignificant, as $g^2 = 10^{-4}$ was used.
It should also be noted that a time step of 0.01 was used in Figure \ref{fig:scaling_params}.
Therefore, the TTS in Figure \ref{fig:scaling_params} is an order of magnitude longer than the results presented in this study, where a time step of 0.125 was used.

\section*{Appendix D: Results and Discussion for CIM-SFC on G-set}
Based on our understanding of CIM-SFC, it is very important that the term $\tanh(c z_i)$ transitions from the ``soft spin" mode where 
$c z_i \approx 0$ and $\tanh(c z_i) \approx c z_i$ to the ``discrete spin" mode where $ |c z_i| >> 0$ and $\tanh(c z_i) \approx \textrm{sign} (c z_i)$.
Therefore, we use the normalizing factor $\xi$ (as defined above) as this ensures that $z_i$ will on average be around $\sqrt{2}$ for a randomly chosen spin configuration,
thus we can use the same value for $c$ in all the cases and get similar results. 
However, this only works for instances such as SK instances where each node has equal connectivity; thus, we can expect $z_i$ to 
have roughly the same range of values for all $i$.

\medskip
On some G-set instances, especially the planar graph instances, some nodes have a much larger degree; thus, $c z_i$ will be too large in some cases and too small in others regardless of the normalizing factor $\xi$ used.
This may be one of the reasons why CIM-SFC struggles on many G-set instances, especially planar graphs. This could also be the reason why dSBM struggles on planar graphs, as dSBM relies on the same normalizing factor to get good results. 
CIM-CAC and CIM-CFC do not need this normalizing factor, as they automatically compensate for different values of $\sum_j J_{ij} \sigma_j$, and this might be why they perform well on planar graphs.

\medskip
Meanwhile, for toroidal graphs, the opposite is true, as $\sum_j J_{ij} \sigma_j$ can only take on five different values for these graphs. 
This could mean that the transition from ``soft spin” to ``discrete spin” is rapid in the case of CIM-SFC; thus, we need to carefully tune the parameters to get good results on these graphs.

\medskip
Although this observation regarding the analog/discrete transition may partially explain the poor results on the G-set, it is not a complete explanation. 
For example, CIM-SFC struggles on some random graphs (such as G9) that do not have the above-mentioned property, as each node has similar connectivity.

\medskip
The results for CIM-SFC on the G-set as well as the parameters used are listed below (not all instances were tested).

\subsection*{Results for CIM-SFC on the G-set}

\begin{center}
\begin{tabular}{| c | c | c |}
\hline
Instance & TTS & $P_s$ \\ \hline
G1 & 28470 & 0.194\\ \hline 
G2 & 1531984 & 0.004\\ \hline 
G3 & 130388 & 0.046\\ \hline 
G4 & 435510 & 0.014\\ \hline 
G5 & 380685 & 0.016\\ \hline 
G6 & 112834 & 0.0202\\ \hline 
G7 & 163316 & 0.014\\ \hline 
G8 & 125360 & 0.0182\\ \hline 
G9 & 4604018 & 0.0005\\ \hline 
G10 & 395845 & 0.0058\\ \hline 
G11 & 195461 & 0.0572\\ \hline 
G12 & 128184 & 0.0859\\ \hline 
G13 & 311368 & 0.0363\\ \hline 
G43 & 1702012 & 0.0134375\\ \hline 
G44 & 1742812 & 0.013125\\ \hline 
G45 & 73671209 & 0.0003125\\ \hline 
G46 & 1927483 & 0.011875\\ \hline 
\end{tabular}
\end{center}
For instances G14–G21 (800 node planar graphs) and G51–G54 (1000 node planar graphs), CIM-SFC shows a success probability of either zero or a very small nonzero value. 
Furthermore, 2000 node instances have not been tested.

\medskip
\subsection*{Parameters for CIM-SFC on G-set}

\subsubsection*{Common parameters}
\begin{center}
\begin{tabular}{c | c}
p & -1.0 $\rightarrow$ 1.0 \\ 
\end{tabular}
\end{center}
\subsubsection*{Parameters selected by problem type}
\begin{center}
\begin{tabular}{| c | c | c | c | c | c | c | c | c | c |}
\hline 
Graph Type & Edge Weight & N & Instance \# & $c$ & $\beta$ & $k$ & N step & $\Delta T$\\ \hline
Random & \{+1\} & 800 &  1-5 & 1.0 $\rightarrow$ 3.0 & 0.3 $\rightarrow$ 0.0 & 0.2 & 2666 & 0.15 \\  \hline
Random & \{+1, -1\} & 800 & 6-10 & 1.0 $\rightarrow$ 3.0 & 0.3 $\rightarrow$ 0.0 & 0.2 & 500 & 0.4 \\  \hline
Toroidal & \{+1, -1\} & 800 &  11-13 & 1.4 & 0.05 $\rightarrow$ 0.0 & 0.32 & 2500 & 0.4\\  \hline
Random & \{+1\} & 1000 &  43-46 & 1.4 $\rightarrow$ 4.2 & 0.2 $\rightarrow$ 0.0 & 0.2 & 5000 & 0.2 \\  \hline

\end{tabular}
\end{center}
The parameters for CIM-SFC are chosen experimentally, and the understanding of how the parameters affect the performance and dynamics is limited. 
Once this system is studied more thoroughly, we will propose a more systematic method of choosing parameters so that good performance can be ensured on many different problem types.

\section*{Appendix E: Similarities and Differences Between CIM and SBM Algorithms}
Using continuous analog dynamics to solve discrete optimization problems is a somewhat new concept, and it is interesting to compare these different approaches.{\scriptsize $^{\cite{Ercsey-Ravasz2011, Leleu2019, Goto2021}}$} 
In this appendix, we will briefly discuss some similarities and differences among the three CIM-inspired algorithms and the SBM algorithms.

\medskip
All four systems discussed in Section 6, namely CIM-CAC, CIM-CAC, CIM-SFC, and dSBM, were originally inspired by the same fundamental principle:\cite{Wang2013, Goto2019}:

\medskip
The function
\setcounter{equation}{0} \renewcommand{\theequation}{D\arabic{equation}}
\begin{equation}
	H(x) = \sum_i \left( \frac{x_i^2}{4} - \frac{1 - p}{2}\right)x_i^2 + c \sum_i \sum_j J_{ij} x_i x_j
\end{equation}
can be used as a continuous approximation of the Ising cost function.

\medskip
In the original CIM algorithm, gradient descent is used to find the local minima of $H$, while $H$ is deformed by increasing $p$. This system has two major drawbacks:\cite{Wang2013}
\begin{enumerate}
	\item local minima are stable;
	\item incorrect mapping of the Ising problem to the cost function owing to amplitude heterogeneity.
\end{enumerate}

All four algorithms discussed in Section 6 can be regarded as modifications of the original CIM algorithm, which aim to overcome these two flaws. {\scriptsize $^{\cite{Leleu2019, Goto2019, Tatsumura2021, Goto2021}}$}
In all these algorithms, the first flaw is addressed by adding new degrees of freedom to the system; hence, there are now $2N$ (instead of only $N$) analog variables for $N$ spins. 
In SBM, this is done by including both a position vector, $x_i$, and a velocity/momentum vector $y_i$, while in the modified CIM algorithms we add the auxiliary variable $e_i$.

\medskip
To address the second flaw, the creators of dSBM added discretization and ``inelastic walls”, whereas in CIM- CFC and CIM-SFC, this discretization is not necessary. Using different mechanisms, all three algorithms ensure that the system only has fixed points at the local minima of the Ising Hamiltonian (during the end
of the trajectory), something which is not true for the original CIM algorithm. Because these systems are fundamentally very similar, it should not be surprising that they achieve similar performance.

\medskip
We also note that for dSBM to achieve good performance, it is necessary to use discretization and inelastic walls, which make the system discontinuous. This is particularly useful for implementation on a digital platform, which prefers discrete processes; however, when implementing these algorithms on an analog physical platform, this is not preferred. 
Meanwhile, in the case of CIM-CAC, CIM-CFC, and CIM-SFC, the system evolves continuously; thus, they are much more suitable for analog implementation, such as the optical CIM architecture proposed in this paper.

\medskip
An interesting difference between the CIM and the original bifurcation machine{\scriptsize $^{\cite{Goto2019}}$}, which was referred to as aSBM in \cite{Goto2021}, is that aSBM is a completely unitary dissipation-less system. 
Because of this, aSBM relies on adiabatic evolution for computation (similar to quantum annealing), unlike the dissipative CIM and other Ising heuristics (such as simulated annealing or breakout local search {\scriptsize $^{\cite{BLS}}$}) ,
which rely on some sort of dissipative relaxation. However, in \cite{Goto2021}, the new SBM algorithms deviate from this concept of adiabatic evolution by adding inelastic walls, thus making the new bifurcation machine a dissipative system in which information is lost over time.
It would be interesting to try to understand whether dissipation is in fact necessary for a system to achieve the high performance of the algorithms discussed in this paper. 
For example, one could modify aSBM in a different way that addresses the problem of amplitude heterogeneity but retains the adiabatic nature. 
Whether this is possible is beyond the scope of this paper.

\section*{Appendix F: Optical Implementation of CIM-SFC}

\begin{figure}[!htb]
	\centering
	\includegraphics[scale=0.12]{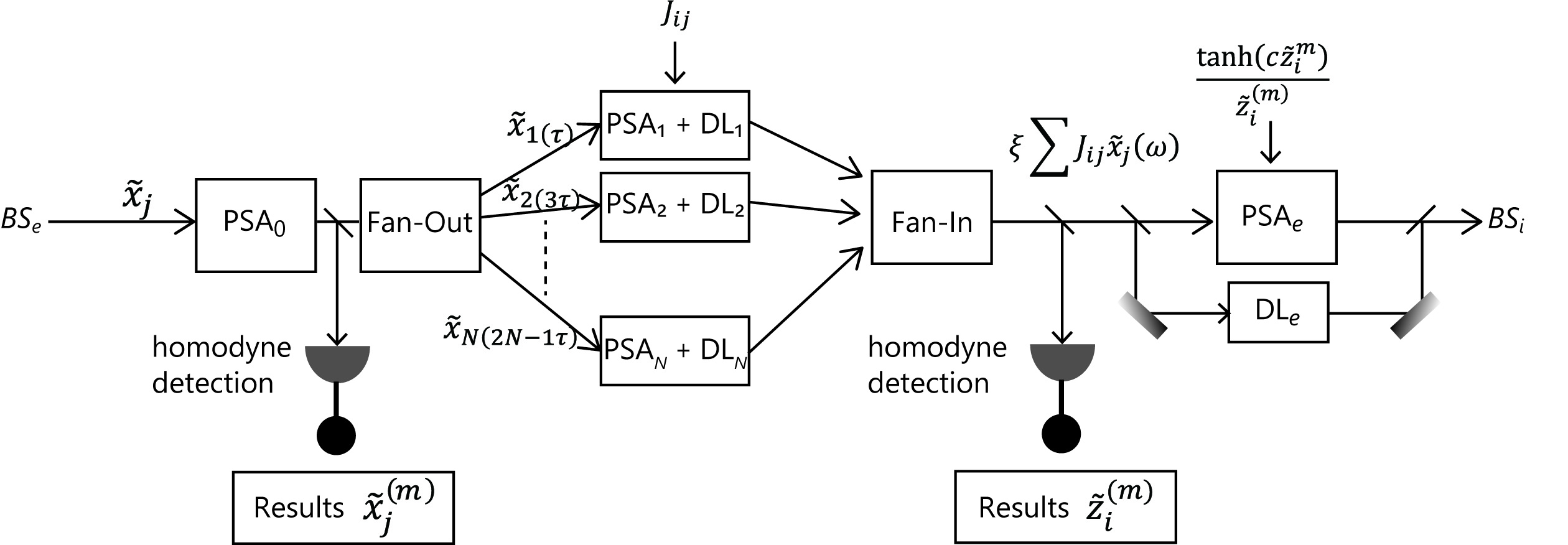}
	\caption{Optical implementation of CIM-SFC.}
	\label{fig:all-optical CIM 2}
\end{figure}

Figure \ref{fig:all-optical CIM 2} shows an optical implementation of CIM-SFC which is similar to that of CIM-CAC and CIM-CFC shown in Figure \ref{fig:all-optical CIM}.
The feedback signal $\tilde{z_i} = \sum_j J_{ij} \tilde{x_j}$ is deamplified (rather than amplified) by PSA$_e$ with attenuation coefficient 
$\frac{\tanh(\tilde{z_i}^{(m)})}{\tilde{z_i}^{(m)}}$, where $\tilde{z_i}^{(m)}$ is an optical homodyne measurement result of $\tilde{z_i}$. This feedback
signal is then injected back into signal pulse $x_i$ in the main cavity through BS$_i$.

\medskip
Part of the fan-in circuit output $\tilde{z_i}$ is delayed by a delay line DL$_e$ with delay time
$N\tau$ and combined with the error pulse $e_i$ (inside the main cavity). This implements the term $e_i - \tilde{z_i}$ in Eq. (\ref{eq:third system 3}).
The term $-\beta (e_i - \tilde{z_i})$ on the right-hand side of Eq. (\ref{eq:third system 3}) is implemented by a phase-sensitive amplifier PSA$_e$ of the main cavity.
This is also a deamplification process. Finally, the error correction signal amplitude $e_i - \tilde{z_i}$ is coupled to the signal pulse $x_i$ inside the 
main cavity with a standard optical delay line.

\medskip
\textbf{Conflict of Interest} \par 
The authors have no confict of interest, financial or otherwise.

\medskip
\textbf{Acknowledgements} \par 
The authors wish to thank R. Hamerly, M. G. Suh, M. Jankowski, E. Ng and Y. Inui for their valuable discussions.

\medskip

%

\bibliographystyle{plain}

\newpage






\end{document}